\documentclass[12pt,journal,draftclsnofoot,oneside,onecolumn]{IEEEtran}

\usepackage{color}
\usepackage[cmex10]{amsmath}
\usepackage{amssymb}
\usepackage{dsfont}
\usepackage{epsfig}
\usepackage{stfloats}
\usepackage{cite}
\usepackage{subfigure}
\usepackage{algorithm}
\usepackage{algorithmic}
\usepackage{setspace}

\newtheorem{lemma}{\it Lemma}
\newtheorem{theorem}{\it Theorem}
\newtheorem{remark}{\it Remark}

\newtheorem{proposition}{\it Proposition}
\begin{document}

\title{Covert Communication in Intelligent Reflecting Surface-Assisted NOMA Systems: Design, Analysis, and Optimization}
\author{Lu Lv,~\IEEEmembership{Member,~IEEE}, Qingqing Wu,~\IEEEmembership{Member,~IEEE},\\ Zan Li,~\IEEEmembership{Senior Member,~IEEE}, Zhiguo Ding,~\IEEEmembership{Fellow,~IEEE},\\ Naofal Al-Dhahir,~\IEEEmembership{Fellow,~IEEE}, and Jian Chen,~\IEEEmembership{Member,~IEEE}
\vspace{-10mm}


\thanks{Lu Lv, Zan Li, and Jian Chen are with the State Key Laboratory of Integrated Services Networks, Xidian University, Xi'an 710071, China (e-mail: \{lulv, zanli\}@xidian.edu.cn; jianchen@mail.xidian.edu.cn).
Qingqing Wu is with the State Key Laboratory of Internet of Things for Smart City, University of Macau, Macau 999078, China (e-mail: qingqingwu@um.edu.mo).
Zhiguo Ding is with the School of Electrical and Electronic Engineering, The University of Manchester, Manchester M13 9PL, U.K. (e-mail: zhiguo.ding@manchester.ac.uk).
Naofal Al-Dhahir is with the Department of Electrical and Computer Engineering, The University of Texas at Dallas, Richardson, TX 75080, USA (e-mail: aldhahir@utdallas.edu).}
}
\maketitle

\begin{abstract}\vspace{-3mm}
  In this paper, we investigate covert communication in an intelligent reflecting surface (IRS)-assisted non-orthogonal multiple access (NOMA) system, where a legitimate transmitter (Alice) applies NOMA for downlink and uplink transmissions with a covert user (Bob) and a public user (Roy) aided by an IRS. Specifically, we propose new IRS-assisted downlink and uplink NOMA schemes to hide the existence of Bob's covert transmission from a warden (Willie), which cost-effectively exploit the phase-shift uncertainty of the IRS and the non-orthogonal signal transmission of Roy as the cover medium without requiring additional uncertainty sources. Assuming the worst-case covert communication scenario where Willie can optimally adjust the detection threshold for his detector, we derive an analytical expression for the minimum average detection error probability of Willie achieved by each of the proposed schemes. To further enhance the covert communication performance, we propose to maximize the covert rates of Bob by jointly optimizing the transmit power and the IRS reflect beamforming, subject to given requirements on the covertness against Willie and the quality-of-service (QoS) at Roy. Simulation results demonstrate the covertness advantage of the proposed schemes and confirm the accuracy of the derived analytical results. Interestingly, it is found that covert communication is impossible without using IRS or NOMA for the considered setup while the proposed schemes can always guarantee positive covert rates.
\end{abstract}

\begin{IEEEkeywords}\vspace{-3mm}
  Intelligent reflecting surface, covert communication, low probability of detection, non-orthogonal multiple access, power allocation, passive beamforming.
\end{IEEEkeywords}
\IEEEpeerreviewmaketitle

\section{Introduction}

Driven by the unprecedented demands for high-data rate applications and ubiquitous wireless services, various advanced wireless technologies including massive multiple-input multiple-output (MIMO) and millimeter wave (mmWave) have been proposed to improve the network performance \cite{JiayiZhang_JSAC2020}. Nevertheless, the benefit generally comes at the expense of high energy consumption and/or hardware cost as well as complexity, due to the use of a huge number of power-hungry active components (i.e., radio-frequency (RF) chains). Against this background, intelligent reflecting surface (IRS), also referred to as reconfigurable intelligent surface, was recently proposed as a cutting-edge technology to achieve spectral- and energy-efficient wireless communications \cite{QingqingWu_CM2020,C.Huang_WCM2020, Y.Liu_ArXiv2020}. Specifically, IRS is a two-dimensional (2D) surface of electromagnetic (EM) material (i.e., metasurface) that consists of a large number of passive and reconfigurable reflecting elements. Each reflecting element can be controlled by a smart controller to adjust the EM properties (e.g., phase and amplitude) of the incoming signals. By smartly controlling all the reflecting elements of the IRS, a desirable radio propagation environment can be established to enhance the data rate or reception reliability \cite{CZhong_JSAC2020,CZhong_WCL2020, C.Huang_JSAC2020}, reduce energy consumption \cite{C.Huang_TWC2019,QingqingWu_TWC2019,QingqingWu_TCOM2020}, extend coverage \cite{Saman_TCOM2020}, and achieve massive connectivity \cite{ZDing_WCL2020, ZengMing_CL2020,TianweiHou_JSAC2020,B.Zheng_TWC2020}.

On the other hand, secrecy and privacy provisioning has become a critical task in developing the sixth-generation (6G) wireless, since the soaring volume of confidential and sensitive data (e.g., financial details, e-health records, and identity authentication) is transmitted over the open wireless medium. This calls for physical layer security, that exploits the intrinsic randomness of noise and fading channels to prevent information leakage \cite{Lu_Network2020, Lu_TIFS2019,Lu_TCOM2020}. Motivated by its capability of reconfiguring wireless channels in a cost-effective manner, IRS is recently integrated with physical layer security to safeguard information transmission. To be specific, with appropriate phase shifting, the IRS reflected signals can be added with the non-reflected signal coherently at the legitimate user but destructively at the eavesdropper, thus significantly improving the secrecy rate. Assuming the knowledge of perfect channel state information (CSI) of the eavesdropper, the secrecy rate maximization and transmit power minimization problems for IRS-assisted multiple-input single-output (MISO) transmission were investigated in \cite{MCiu_WCL2019,ZhengChu_WCL2020}, respectively. After that, attention was shifted to an IRS-assisted MIMO scenario to maximize the secrecy rate in \cite{LimengDong_TWC2020}. In the cases with and without eavesdropper's CSI, the authors of \cite{XinrongGuan_WCL2020,HuimingWang_SPL2020} showed that incorporating artificial noise in transmit beamforming along with IRS reflect beamforming is helpful to increase the secrecy rate. In \cite{Lu_CL2021}, a new information jamming scheme via IRS was proposed to guarantee secrecy for two-way communications.

However, in certain circumstances, protecting the content of communications using existing physical layer security techniques is far from sufficient, and the communication itself is often required to hide from being detected \cite{Bash_CM2015}. For example, in tactical intelligent networks, military operations desire to shield themselves from the adversary. Alternatively, in financial institution networks, an entity hopes to protect its own secret activities from being monitored by an authoritarian government. Thus, secure communication systems should also provide stealth or low probability of detection, and catering to such security concern motivates the recent advances in covert communication. Theoretically, covert communication aims to explore the fundamental limits of hiding the amount of wireless information that can be covertly transmitted from a legitimate transmitter to its legitimate receiver, subject to a negligible probability of being detected by a warden \cite{Bash_CM2015, YanShihao_WCM2019}. The work of \cite{BiaoHe_CL2017} pointed out that a positive covert rate can be achieved when the warden does not exactly know its received noise power. For the scenario where the warden has uncertainty about the aggregate received interference, the studies in \cite{BiaoHe_TWC2018,ZhengTX_TWC2019} investigated the maximum covert throughput and the covert outage probability in random wireless networks. Covert communication with delay constraints was studied in \cite{YanShihao_TIFS2019}, where the authors showed that the incorporation of finite block length and the use of random transmit power can effectively induce confusion to the warden. In \cite{HuJingsong_TVT2019}, both conventional and truncated channel inversion power control strategies were proposed to prevent the warden from knowing the existence of the legitimate transmitter and thus guarantee covertness. Furthermore, it was found in \cite{HuJingsong_TWC2018,Y.Jiang_TVT2020,YangWeiwei_WCL2020} that covert communication with a positive covert rate can be achieved by embedding the covert signal into a superimposed signal and making use of the non-covert transmissions of a greedy relay \cite{HuJingsong_TWC2018}, a cellular user \cite{Y.Jiang_TVT2020}, and a non-orthogonal multiple access (NOMA) weak user \cite{YangWeiwei_WCL2020}, as the shield.

It is worth noting that the reviewed covert strategies in \cite{BiaoHe_CL2017, BiaoHe_TWC2018,ZhengTX_TWC2019,YanShihao_TIFS2019,HuJingsong_TVT2019,HuJingsong_TWC2018, Y.Jiang_TVT2020,YangWeiwei_WCL2020} may degrade the communication performance at legitimate users to a certain extent, due to the resource-consuming issue and the stringent covertness constraint \cite{LuXiao_Network2020}. To address this dilemma, several recent studies have attempted to exploit the IRS to facilitate covert wireless communication via simultaneously improving the received signal quality at the legitimate user and weakening the signal strength at the warden. In \cite{LuXiao_Network2020}, by considering the noise uncertainty at the warden as the cover medium, the authors showed that a much higher covert rate can be achieved by using an IRS compared to that without an IRS. The work of \cite{LuXiao_Network2020} was then extended to a more general system setup with both single antenna and multiple antennas at the legitimate transmitter in \cite{JiangboSi_ArXiv2020}, where the impact of different CSI availabilities of the warden on the covert rate performance is evaluated. More lately, the authors in \cite{XiaoboZhou_ArXiv2020} investigated the joint design of the transmit power and the IRS reflection coefficients to maximize the received signal power of the covert receiver, under the assumption of a finite number of channel uses.

As for the aforementioned research efforts, we make two important observations below.
\begin{itemize}
  \item In many existing works (see, e.g., \cite{HuJingsong_TWC2018,Y.Jiang_TVT2020, YangWeiwei_WCL2020}) on using public actions as the cover medium for the covert action, it is assumed that different codebooks \cite{HuJingsong_TWC2018} or random transmit power \cite{Y.Jiang_TVT2020, YangWeiwei_WCL2020} is applied at the legitimate transmitter. However, we note that: i) the assumption of different codebooks at the legitimate transmitter may not be feasible for all possible scenarios; and ii) as indicated in \cite{YanShihao_TIFS2019}, with random transmit power, both the number of transmit power levels and codebooks for transmission rate should approach infinity, which is often difficult to achieve in practical wireless networks.

  \item Initial studies (see, e.g., \cite{LuXiao_Network2020,JiangboSi_ArXiv2020, XiaoboZhou_ArXiv2020}) proposed to use the IRS for covert communication, but extra uncertainty sources, e.g., noise uncertainty at the warden \cite{LuXiao_Network2020,JiangboSi_ArXiv2020}, or finite block length requirement \cite{XiaoboZhou_ArXiv2020} (that is limited to some specified application scenarios) is needed. In fact, they all neglect a useful source of uncertainty that exists inherently in the considered IRS-assisted covert communication system. Specifically, the phase shifts of the IRS can be designed to induce a deliberate confusion and degrade the signal detection performance at the warden, while at the same time enhancing the signal reception quality of the legitimate receiver to benefit covertness. However, such a novel design has not yet been considered for covert communication, to our best knowledge.
\end{itemize}

Motivated by the above observations, as the first work, we study both downlink and uplink covert communications in an IRS-assisted NOMA system, where a legitimate transmitter applies NOMA to communicate with a covert user and a public user against a warden. Our goal is to hide the communication between the legitimate transmitter and the covert user by making full use of the uncertainty inherent in the wireless system environment. The main contributions of this paper are summarized as follows.
\begin{enumerate}
  \item We propose novel IRS-assisted downlink and uplink NOMA schemes to achieve covert wireless communications. The phase-shift uncertainty of the IRS and the non-orthogonal signal transmission of the public user are jointly exploited as the new cover medium to shield the signal transmission of the covert user. As such, the proposed schemes do not require any other uncertainty sources such as the transmitter's random transmit power or the warden's noise uncertainty, thus rendering them generally simpler and more cost-effective than existing covert communication approaches.


  \item Under the worst-case scenario from the perspective of the covert communication where the warden can optimally choose its detection threshold, we derive an analytical closed-form expression for the minimum average detection error probability of the warden achieved by each of the proposed schemes.

  \item For each of the proposed schemes, a joint optimization framework of the transmit power allocation and the IRS reflect beamforming is formulated to enhance the performance of the covert communication. To deal with the formulated non-convex optimization problems, we further develop efficient algorithms based on the alternating optimization to optimize the power allocation and reflection coefficients alternatively. In particular, at each iteration, the optimal power allocation solutions for given IRS phase shifts are derived in closed form, and the optimal reflection coefficients for given power allocation are obtained relying on the semidefinite relaxation (SDR) technique.

  \item Through analytical and numerical results, we obtain various useful insights: i) our schemes can always guarantee positive covert rates with non-zero transmit power; ii) increasing the transmit power for the public user and the number of reflecting elements at the IRS are both helpful to degrade warden's detection performance. Particularly, as the transmit power for the public user grows large, the minimum average detection error probability of the warden approaches one, implying that the warden's detection seems like a random guess; and iii) it is impossible to achieve covert communications without using an IRS or NOMA in the considered system, thus validating the effectiveness of the proposed schemes.
\end{enumerate}

The rest of the paper is organized as follows. Section~\ref{sec:model} introduces the system model under investigation. Sections~\ref{sec:downlink} and \ref{sec:uplink} provide the transmission design, performance analysis, and power/beamforming optimization framework for the IRS-assisted downlink and uplink NOMA schemes, respectively, to achieve covert wireless communications. Simulation results are presented in Section~\ref{sec:simulation}. Finally, we conclude the paper in Section~\ref{sec:conclusion}.

{\it Notations:} In this paper, scalars are denoted by italic letters, vectors and matrices are denoted by bold-face lower-case and upper-case letters, respectively. $\Pr(\cdot)$ denotes probability, $f_X(\cdot)$ denotes the probability density function (PDF), and $\arg(x)$ denote the phase of the complex-valued random variable (RV) $x$. The rank and the trace of matrix $\mathbf{S}$ are denoted by $\mathrm{rank}(\mathbf{S})$ and $\mathrm{tr}(\mathbf{S})$. Furthermore, $\mathbf{S}\succeq\mathbf{0}$ implies that $\mathbf{S}$ is a positive semidefinite matrix.

\section{System Model}
\label{sec:model}

\begin{figure}[t]
  \centering
  \includegraphics[width=5.0in]{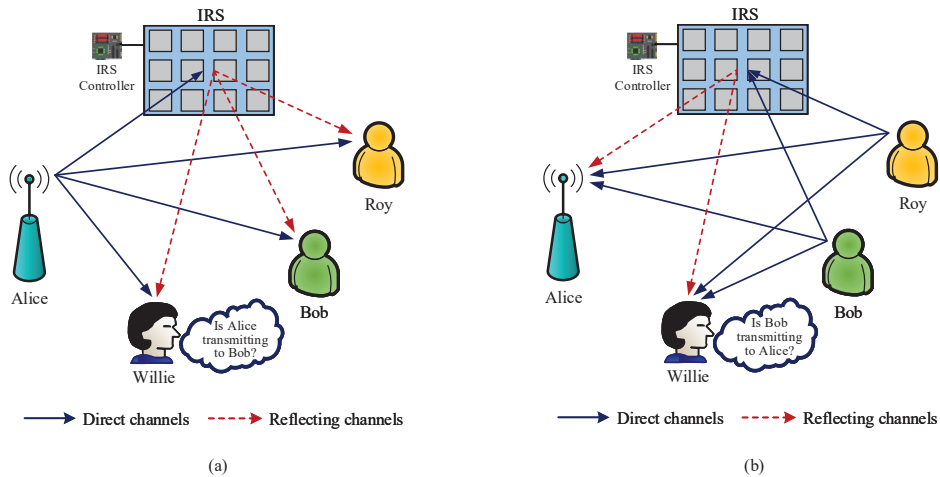}\vspace{-2mm}\\
  \caption{Covert communication in an IRS-assisted NOMA system. (a) Downlink scenario. (b) Uplink scenario.}
  \label{fig:system-model}\vspace{-3mm}
\end{figure}

We consider covert communication in an IRS-assisted NOMA system, which consists of a legitimate transmitter (called Alice), a covert user (called Bob), a public user (called Roy), an IRS, and a warden (called Willie), as shown in Fig.~\ref{fig:system-model}. The IRS is deployed to assist in the downlink and uplink public/covert signal transmissions between Alice and Roy/Bob, while guaranteeing a low probability of being detected by Willie. Each node is equipped with a single antenna and operates in a half-duplex mode, which means that signal transmission and reception cannot be carried out on the same time/frequency resource block. The IRS has $N$ low-cost reconfigurable reflecting elements. Each of the elements can reflect a phase shifted version of the incident signal independently, aiming at improving the signal reception qualities at Roy and Bob (i.e., in downlink) or Alice (i.e., in uplink) as well as creating uncertainty at Willie for covertness enhancement. Due to the severe path loss, we assume that the powers of the signals that are reflected by the IRS two or more times are sufficiently weak, so that they can be ignored \cite{QingqingWu_CM2020,QingqingWu_TWC2019}. Furthermore, the IRS is connected to a smart controller, such as field-programmable gate array (FPGA), that is in charge of the reflection reconfiguration and assists in the channel estimation of the IRS-involved links \cite{QingqingWu_CM2020,Y.Liu_ArXiv2020, B.Zheng_TWC2020}.

All the wireless channels are assumed to experience a quasi-static block fading along with a distance-based path loss. The fading coefficient vectors between Alice/Bob/Roy/Willie and IRS are denoted by $\mathbf{h}_a$, $\mathbf{h}_b$, $\mathbf{h}_r$, and $\mathbf{h}_w$, with the corresponding distances being $d_a$, $d_b$, $d_r$, and $d_w$. The fading coefficients between Alice and Bob/Roy/Willie are denoted by $h_{ab}$, $h_{ar}$, and $h_{aw}$, with the corresponding distances being $d_{ab}$, $d_{ar}$, and $d_{aw}$. The fading coefficients between Bob/Roy and Willie are denoted by $h_{bw}$ and $h_{rw}$, with the corresponding distances being $d_{bw}$ and $d_{rw}$. We assume a time-division duplexing (TDD) protocol for downlink and uplink transmissions, such that channel reciprocity holds. Each entry of $\mathbf{h}_a$, $\mathbf{h}_b$, $\mathbf{h}_r$, and $\mathbf{h}_w$, as well as $h_{ab}$, $h_{ar}$, $h_{aw}$, $h_{bw}$, and $h_{rw}$, are independent and identically distributed with zero mean and unit variance. Furthermore, each receiving node is corrupted by additive white Gaussian noise (AWGN) with zero mean and variance of $\sigma_0^2$.

The CSI availability is discussed next. For the downlink scenario, we assume that: 1) Alice knows the instantaneous CSI of the direct Alice-user links and the reflected Alice-IRS-user links, but only knows the statistical CSI of the direct Alice-Willie link and the IRS-Willie link. This is reasonable because Willie is usually an external warden and tries to hide its existence from the legitimate system, and it is difficult for Alice to know its instantaneous CSI. 2) Willie possesses the instantaneous CSI of the direct Alice-Willie link and the reflected Alice-IRS-Willie link, which is the worst-case scenario from the perspective of covert communication design. Similarly, for the uplink scenario, we assume that: 1) Roy knows the instantaneous CSI of the direct Roy-Alice link and the reflected Roy-IRS-Alice link but only the statistical CSI of the Roy-Willie link and the IRS-Willie link. In addition, Bob knows instantaneous CSI of the direct Bob-Alice link and the reflected Bob-IRS-Alice link but only the statistical CSI of the Bob-Willie link and the IRS-Willie link. 2) Willie possesses the instantaneous CSI of the direct Roy/Bob-Willie links and the reflected Roy/Bob-IRS-Willie links.

\section{Covert Communication in IRS-Assisted Downlink NOMA}
\label{sec:downlink}

This section first proposes a novel IRS-assisted downlink NOMA scheme to achieve covert wireless communication. Then, an analytical closed-form expression for the minimum average detection error probability is derived by assuming that Willie can optimally choose his detection threshold. Finally, an efficient alternating optimization algorithm is developed to solve the joint transmit power and passive beamforming optimization problem to maximize the covert rate.

\subsection{IRS-Assisted Downlink NOMA Scheme}

As shown in Fig.~\ref{fig:system-model}(a), by applying the downlink NOMA principle, Alice transmits a public signal $s_r(k)$ and a covert signal $s_b(k)$ to Roy and Bob, respectively, where $k=1,\dots,K$ denotes the index of each signal sample, $K$ is the total number of signal samples in a communication slot, and $s_r(k)$ and $s_b(k)$ are Gaussian input signals with zero mean and unit variance. Here, Roy's public signal $s_r(k)$ is exploited as a cover for Bob's covert signal $s_b(k)$. Thus, the received signals at Bob and Roy are given, respectively, by
\begin{align}
  y_b(k)&=\bigg(\frac{h_{ab}}{\sqrt{L(d_{ab})}}+ \frac{\mathbf{h}_a^H\mathbf{\Theta}\mathbf{h}_b}{\sqrt{L(d_a)L(d_b)}}\bigg) \Big(\sqrt{P_r}s_r(k)+\sqrt{P_b}s_b(k)\Big)+z_b(k), \label{DL-yb}\\
  y_r(k)&=\bigg(\frac{h_{ar}}{\sqrt{L(d_{ar})}} +\frac{\mathbf{h}_a^H\mathbf{\Theta}\mathbf{h}_r}{\sqrt{L(d_a) L(d_r)}}\bigg)\Big(\sqrt{P_r}s_r(k)+\sqrt{P_b}s_b(k)\Big)+z_r(k), \label{DL-yr}
\end{align}
where $\mathbf{\Theta}=\text{diag}(e^{j\theta_1},\dots,e^{j\theta_N})\in\mathbb{C}^{N\times N}$ denotes the IRS diagonal reflection-coefficients matrix with $\theta_n\in[0,2\pi)$ being the phase shift of the $n$th IRS element. Since the instantaneous CSI of Willie is unavailable to Alice, the optimal $\mathbf{\Theta}$ should be selected based on the CSI of the Alice-Bob/Roy and Alice-IRS-Bob/Roy links, which is discussed in detail in Section III-C. $P_r$ and $P_b$ denote the average transmit power of $s_r(k)$ and $s_b(k)$, respectively. $z_r(k)$ and $z_b(k)$ denote the AWGNs at Roy and Bob, respectively, and $L(d)$ denotes the effective path loss function \cite{ZDing_WCL2020}. To facilitate the IRS-assisted downlink NOMA scheme, the users are ordered according to their composite channel (which includes the direct Alice-user link and reflected Alice-IRS-user link) gains, which has been widely adopted in the literature on IRS-assisted NOMA \cite{ZDing_WCL2020,ZengMing_CL2020,TianweiHou_JSAC2020}. Without loss of generality, we assume that the composite channel gains of the users are ordered as $|g_{ab}|^2\geq|g_{ar}|^2$, where $g_{ab}=\frac{h_{ab}}{\sqrt{L(d_{ab})}}+ \frac{\mathbf{h}_a^H\mathbf{\Theta}\mathbf{h}_b}{\sqrt{L(d_a) L(d_b)}}$ is the composite channel of Bob and $g_{ar}=\frac{h_{ar}}{\sqrt{L(d_{ar})}} +\frac{\mathbf{h}_a^H\mathbf{\Theta}\mathbf{h}_r}{\sqrt{L(d_a)L(d_r)}}$ is the composite channel of Roy. It is important to point out that with the above ordered channels, the SIC ordering at Bob should start from $s_r(k)$ first and then move towards $s_b(k)$, and thus, more transmit power is allocated to Roy, i.e., $P_r\geq P_b$, to guarantee successful SIC \cite{ZDing_WCL2020}. This setting is beneficial to the covertness of Bob, since a higher transmit power of Roy can better protect Bob's covert communication. Furthermore, this SIC order does not affect the performance of Willie, due to the fact that Willie only uses the energy detection to detect the existence of Bob's transmission rather than decoding the signals. Therefore, the achievable rates of Bob to sequentially decode $s_r(k)$ and $s_b(k)$ are given by
\begin{align}
  R_{b,s_r(k)}&=\log\bigg(1+\frac{P_r|g_{ab}|^2}{P_b|g_{ab}|^2+\sigma_0^2}\bigg), \label{DL-rate-Bob-sr}\\
  R_{b,s_b(k)}&=\log\bigg(1+\frac{P_b|g_{ab}|^2}{\sigma_0^2}\bigg). \label{DL-rate-Bob-sb}
\end{align}
Roy directly decodes $s_r(k)$ by treating $s_b(k)$ as noise, yielding the achievable rate as
\begin{equation}
  R_{r,s_r(k)}=\log\bigg(1+\frac{P_r|g_{ar}|^2}{P_b|g_{ar}|^2+\sigma_0^2}\bigg).
\end{equation}


On the other hand, Willie tries to judge whether Alice is transmitting a covert signal to Bob or not by carrying out the Neyman-Pearson test based on his received signal sequence $y_w(k)$ for $k=1,\dots,K$ \cite{ZhengTX_TWC2019}. As a result, Willie faces a binary detection problem: 1) the null hypothesis $\mathcal{H}_0$ which indicates Alice is not transmitting to Bob, and 2) the alternative hypothesis $\mathcal{H}_1$ which indicates an ongoing covert transmission from Alice to Bob. Then, the received signals at Willie under these two hypotheses are given, respectively, by
\begin{align}
  \mathcal{H}_0:\ y_w(k)&=\bigg(\frac{h_{aw}}{\sqrt{L(d_{aw})}}+ \frac{\mathbf{h}_a^H\mathbf{\Theta}\mathbf{h}_w}{\sqrt{L(d_a)L(d_w)}}\bigg) \sqrt{P_r}s_r(k)+z_w(k), \label{DL-yw0}\\
  \mathcal{H}_1:\ y_w(k)&=\bigg(\frac{h_{aw}}{\sqrt{L(d_{aw})}}+ \frac{\mathbf{h}_a^H\mathbf{\Theta}\mathbf{h}_w}{\sqrt{L(d_a)L(d_w)}}\bigg) \Big(\sqrt{P_r}s_r(k)+\sqrt{P_b}s_b(k)\Big)+z_w(k), \label{DL-yw1}
\end{align}
where $z_w(k)$ denotes the AWGN at Willie. Based on \eqref{DL-yw0} and \eqref{DL-yw1}, Willie adopts a radiometer for the binary detection. Using the average received power at Willie (i.e., $P_w=\frac1K\sum_{k=1}^K|y_w(k)|^2$) as the test statistic, the decision rule is given by
\begin{equation}
\label{test-rule}
  P_w\underset{\mathcal{D}_0}{\overset{\mathcal{D}_1}{\gtrless}}\tau_\text{dl},
\end{equation}
where $\tau_\text{dl}>0$ is the detection threshold of Willie's detector, $\mathcal{D}_1$ and $\mathcal{D}_0$ are the binary decisions in favor of $\mathcal{H}_1$ and $\mathcal{H}_0$, respectively. Similar to \cite{BiaoHe_CL2017, BiaoHe_TWC2018,ZhengTX_TWC2019}, we assume that Willie uses an infinite number of signal samples for the binary detection, i.e., $K\rightarrow\infty$, such that the uncertainties of the transmitted signals and the received AWGNs will vanish. Accordingly, the average received power at Willie of the proposed IRS-assisted downlink NOMA scheme is obtained as
\begin{equation}
\label{Pw}
  P_w=\begin{cases}
    \Big(\frac{|h_{aw}|^2}{L(d_{aw})}+\frac{|\mathbf{h}_a^H\mathbf{\Theta}\mathbf{h}_w|^2} {L(d_a)L(d_w)}\Big)P_r+\sigma_0^2, & \mathcal{H}_0,\\
    \Big(\frac{|h_{aw}|^2}{L(d_{aw})} +\frac{|\mathbf{h}_a^H\mathbf{\Theta}\mathbf{h}_w|^2}{L(d_a) L(d_w)}\Big)\big(P_r+P_b\big)+\sigma_0^2, & \mathcal{H}_1.
  \end{cases}
\end{equation}
The performance of Willie's hypothesis test can be measured by the detection error probability, which is defined as
\begin{equation}
\label{DEP-definition}
  \xi_{w,\text{dl}}\triangleq\mathbb{P}_\text{FA}+\mathbb{P}_\text{MD},
\end{equation}
where $\mathbb{P}_\text{FA}=\mathbb{P}(\mathcal{D}_1|\mathcal{H}_0)$ denotes the false alarm probability, $\mathbb{P}_\text{MD}=\mathbb{P}(\mathcal{D}_0|\mathcal{H}_1)$ denotes the miss detection probability, and $0\leq\xi_{w,\text{dl}}\leq1$. To be specific, $\xi_{w,\text{dl}}=0$ implies that Willie can perfectly detect the covert signal without error, while $\xi_{w,\text{dl}}=1$ implies that Willie fails to detect the covert signal and his behavior is like a random guess.

\begin{remark}
  To guarantee Bob's covert communication, the key idea is to artificially create the phase-shift uncertainty at the IRS. Since the optimal $\mathbf{\Theta}$ is designed according to the CSI of the Alice-Bob/Roy and Alice-IRS-Bob/Roy links, this makes $|\mathbf{h}_a^H\mathbf{\Theta}\mathbf{h}_w|^2$ in \eqref{Pw} a RV at Willie and only the distribution of $|\mathbf{h}_a^H\mathbf{\Theta}\mathbf{h}_w|^2$ is known to him. In this way, when Willie measures his received power, he cannot tell whether the received power change is due to the public signal transmission or the covert signal transmission, thus ensuring a low probability of detection with respect to Bob's covert communication. As a result, we conclude that the proposed IRS-assisted downlink NOMA scheme is cost-effective and easy to implement, in the sense that the uncertainty exists inherently in Roy's non-orthogonal transmission and the phase-shift uncertainty of the IRS, without requiring any extra uncertainty sources, e.g., random transmit power at the legitimate user \cite{Y.Jiang_TVT2020,YangWeiwei_WCL2020} or uncertain noise power at the warden \cite{LuXiao_Network2020,JiangboSi_ArXiv2020}, which are usually difficult to implement in practice.
\end{remark}

\begin{remark}
  Consider an IRS-assisted orthogonal multiple access (OMA) scenario where the signals for Roy and Bob are transmitted separately in two orthogonal time slots, the covertness of Bob cannot be achieved for the considered setup. Since without the instantaneous CSI of Willie, it is impossible to design the IRS to help neutralize the signals received by Willie completely. Thus, Willie can exactly measure his non-zero received power of Bob's covert communication. Furthermore, instead of using IRS, we consider a NOMA full-duplex relay scenario where the IRS is replaced by a full-duplex relay with constant transmit power. In this case, Bob's covert communication cannot be achieved, since Willie can easily raise an alarm if additional transmit power for Bob's covert signal is received.
\end{remark}

\subsection{Detection Error Probability of Willie}


For ease of notation, we denote $\delta_N=\mathbf{h}_a^H\mathbf{\Theta}\mathbf{h}_w $, which can be re-expressed as
\begin{equation}
  \delta_N=\sum_{n=1}^N|h_{an}||h_{wn}|e^{-j\psi_n},
\end{equation}
where $h_{an}$ and $h_{wn}$ are the fading coefficients from Alice and Willie to the $n$th element of the IRS, and $\psi_n=\theta_n^\ast+\arg(h_{an})+\arg(h_{wn})$. Specifically, $\theta_n^\ast$ denotes the optimal phase shift of the IRS in the downlink, which is a function of $\arg(h_{an})$, $\arg(h_{bn})$, $\arg(h_{rn})$, $\arg(h_{ab})$, and $\arg(h_{ar})$. Before deriving the distribution of $\delta_N$, we first characterize the distribution of $\psi_n$ in the following lemma.

\begin{lemma}
  The phase value $\psi_n$ follows an independent and uniform distribution on $[0,2\pi)$, namely, $f_{\psi_n}(x)=\frac1{2\pi}$ with $x\in[0,2\pi)$.
\end{lemma}
\begin{IEEEproof}
  Please refer to Appendix A.
\end{IEEEproof}

In fact, it is challenging to obtain the exact distribution of $\delta_N$ since $\delta_N$ is a sum of complex-valued RVs with correlated imaginary and real coefficients. However, when $N$ is sufficiently large, using Lemma 1 and \cite[Lemma~2]{ZDing_WCL2020}, the CDF of $\delta_N$ can be approximated by a complex Gaussian RV with zero mean and variance equal to $N$, i.e., $\delta_N\sim\mathcal{CN}(0,N)$. Moreover, as validated by the numerical results in \cite{ZDing_WCL2020}, such an approximation is also very tight with relatively small $N$.

Accordingly, the false alarm probability of the proposed IRS-assisted downlink NOMA scheme can be derived as
\begin{align}
\label{FA-downlink}
  \mathbb{P}_\text{FA}&=\Pr\bigg(\frac{P_r|h_{aw}|^2}{L(d_{aw})}+\frac{P_r\delta_N^2} {L(d_a)L(d_w)}+\sigma_0^2>\tau_\text{dl}\bigg) \nonumber\\
  &=\begin{cases}
    \underbrace{e^{\frac{P_r|h_{aw}|^2/L(d_{aw})+\sigma_0^2-\tau_\text{dl}} {P_rN/(L(d_a)L(d_w))}}}_{\nu_1}, &\text{if}\ \tau_\text{dl}>\sigma_0^2+\frac{P_r|h_{aw}|^2}{L(d_{aw})}, \\
    1, &\text{otherwise}.
  \end{cases}
\end{align}
Moreover, the miss detection probability of the proposed IRS-assisted downlink NOMA scheme is computed by
\begin{align}
\label{MD-downlink}
  \mathbb{P}_\text{MD}&=\Pr\bigg(\frac{(P_r+P_b)|h_{aw}|^2}{L(d_{aw})} +\frac{(P_r+P_b)\delta_N^2}{L(d_a)L(d_w)}+\sigma_0^2<\tau_\text{dl}\bigg)\nonumber\\
  &=\begin{cases}
    1-\underbrace{e^{\frac{(P_r+P_b)|h_{aw}|^2/L(d_{aw})+\sigma_0^2-\tau_\text{dl}} {(P_r+P_b)N/ (L(d_a)L(d_w))}}}_{\nu_2}, & \text{if}\ \tau_\text{dl}>\sigma_0^2+\frac{(P_r+P_b)|h_{aw}|^2}{L(d_{aw})},\\
    0, & \text{otherwise}.
  \end{cases}
\end{align}
Substituting \eqref{FA-downlink} and \eqref{MD-downlink} into \eqref{DEP-definition}, the detection error probability of the proposed IRS-assisted downlink NOMA scheme is obtained as
\begin{equation}
\label{DEP-downlink}
  \xi_{w,\text{dl}}=\begin{cases}
    1, & \text{if}\ \tau_\text{dl}<\sigma_0^2+\frac{P_r|h_{aw}|^2}{L(d_{aw})},\\
    \nu_1, & \text{if}\ \sigma_0^2+\frac{P_r|h_{aw}|^2}{L(d_{aw})}\leq\tau_\text{dl}\leq \sigma_0^2+\frac{(P_r+P_b)|h_{aw}|^2}{L(d_{aw})},\\
    1+\nu_1-\nu_2, & \text{if}\ \tau_\text{dl}>\sigma_0^2+\frac{(P_r+P_b)|h_{aw}|^2}{L(d_{aw})}.
  \end{cases}
\end{equation}
It is important to point out that the detection error probability in \eqref{DEP-downlink} is derived by assuming an arbitrary detection threshold $\tau_\text{dl}$. From a worst-case perspective of covert communication, Willie can optimally choose $\tau_\text{dl}$ to achieve the minimum detection error probability, i.e., $\min_{\tau_\text{dl}}\xi_{w,\text{dl}}$. The optimal solution of $\tau_\text{dl}$ is provided in the following theorem.
\begin{theorem}
  The optimal detection threshold of Willie to minimize the detection error probability of the IRS-assisted downlink NOMA scheme is obtained as
  \begin{equation}
  \label{downlink-optimal-threshold}
    \tau_\text{dl}^\ast=\begin{cases}
      \sigma_0^2+\frac{(P_r+P_b)|h_{aw}|^2}{L(d_{aw})}, & \text{if}\ |h_{aw}|^2\geq \frac{P_rN}{P_b\phi_1}\ln\big(\frac{P_r+P_b}{P_r}\big),\\
      \sigma_0^2+\frac{P_r(P_r+P_b)N}{P_bL(d_a)L(d_w)} \ln\big(\frac{P_r+P_b}{P_r}\big), & \text{otherwise},
    \end{cases}
  \end{equation}
  where $\phi_1=\frac{L(d_a)L(d_w)}{L(d_{aw})}$.
\end{theorem}
\begin{IEEEproof}
  Please refer to Appendix B.
\end{IEEEproof}

Then, by substituting \eqref{downlink-optimal-threshold} into \eqref{DEP-downlink} and applying some algebraic manipulations, the minimum detection error probability of the IRS-assisted downlink NOMA scheme is expressed as
\begin{equation}
\label{minDEP-downlink}
  \xi_{w,\text{dl}}^\ast=\begin{cases}
    e^{-\frac{P_b\phi_1|h_{aw}|^2}{P_rN}}, & \text{if}\ |h_{aw}|^2\geq \frac{P_rN}{P_b\phi_1}\ln\big(\frac{P_r+P_b}{P_r}\big),\\
    1-\frac{P_b}{P_r}\big(\frac{P_r+P_b}{P_r}\big)^{-\frac{P_r+P_b}{P_b}} e^{\frac{\phi_1|h_{aw}|^2}{N}}, & \text{otherwise}.
  \end{cases}
\end{equation}

Recall that Alice does not know the instantaneous CSI of $h_{aw}$. Hence, we adopt the minimum average detection error probability over all channel realizations of $|h_{aw}|^2$ as the performance metric of covertness. The following theorem provides a closed-form expression for the minimum average detection error probability.
\begin{theorem}
  The minimum average detection error probability of Willie achieved by the IRS-assisted downlink NOMA scheme is given by
  \begin{align}
  \label{minDEP-downlink-closed}
    \bar{\xi}_{w,\text{dl}}^\ast&=\frac{\big(\frac{P_r+P_b}{P_r}\big)^ {-(1+\frac{P_rN}{P_b\phi_1})}}{P_b\phi_1(P_rN)^{-1}+1}+ 1-\Big(\frac{P_r+P_b}{P_r}\Big)^{-\frac{P_rN}{P_b\phi_1}} \nonumber\\
    &\quad-\frac{\frac{P_b}{P_r}\big(\frac{P_r+P_b}{P_r}\big) ^{-\frac{P_r+P_b}{P_b}}}{\phi_1 N^{-1}-1} \bigg[\Big(\frac{P_r+P_b}{P_r}\Big)^{\frac{P_r}{P_b}(1-\frac{N}{\phi_1})}-1\bigg].
  \end{align}
\end{theorem}
\begin{IEEEproof}
  According to the Total Probability Theorem, the minimum average detection error probability of the IRS-assisted downlink NOMA scheme can be expressed as
  \begin{align}
  \label{expectation-downlink}
    \bar{\xi}_{w,\text{dl}}^\ast&=\int_{\frac{P_rN}{P_b\phi_1} \ln\big(\frac{P_r+P_b}{P_r}\big)}^\infty e^{-\frac{P_b\phi_1x}{P_rN}}f_{|h_{aw}|^2}(x)dx \nonumber\\
    &\quad+\int_0^{\frac{P_rN}{P_b\phi_1}\ln\big(\frac{P_r+P_b}{P_r}\big)}\bigg[1- \frac{P_b}{P_r}\Big(\frac{P_r+P_b}{P_r}\Big)^{-\frac{P_r+P_b}{P_b}}e^{\frac{\phi_1 x}{N}}\bigg]f_{|h_{aw}|^2}(x)dx.
  \end{align}
  By calculating the integral in \eqref{expectation-downlink}, we can readily obtain \eqref{minDEP-downlink-closed}, which completes the proof.
\end{IEEEproof}

\begin{remark}
  Based on \eqref{minDEP-downlink-closed}, we reveal the following useful insights: 1) $\bar{\xi}_{w,\text{dl}}^\ast\rightarrow0$, when $P_r$ is finite but $P_b\rightarrow\infty$; and 2) $\bar{\xi}_{w,\text{dl}}^\ast\rightarrow1$, when $P_b$ is finite but $P_r\rightarrow\infty$. The above results can be readily proved using the L'Hopital's Rule in the large $P_b$ or $P_r$ regime, and thus omitted here for simplicity.
\end{remark}


%
%

\subsection{Joint Power and Beamforming Optimization}

To improve the covert performance of the IRS-assisted downlink NOMA system, we aim to maximize the covert rate of Bob by jointly optimizing the transmit power of Alice and the reflect beamforming of the IRS, subject to the total transmit power constraint at Alice, the quality-of-service (QoS) constraint at Roy, and the minimum average detection error probability constraint at Willie. Accordingly, the optimization problem can be formulated as
\begin{subequations}
  \begin{align}
    \text{(P1):}\ &\max_{P_r,P_b,\mathbf{\Theta}}\ R_{b,s_b(k)} \label{DOP-1a}\\
    &\text{s.t.}\,\,  P_r+P_b\leq P_a^{\max},\ P_r\geq P_b, \label{DOP-1b}\\
    &\quad\ R_{r,s_r(k)}\leq R_{b,s_r(k)}, \label{DOP-1c}\\
    &\quad\ R_{r,s_r(k)}\geq R_{r,s_r(k)}^{\min}, \label{DOP-1d}\\
    &\quad\ \bar{\xi}_{w,\text{dl}}^\ast\geq1-\varepsilon,\ \varepsilon\in[0,1], \label{DOP-1e}\\
    &\quad\ 0\leq\theta_n\leq2\pi,\ n=1,\dots,N. \label{DOP-1f}
  \end{align}
\end{subequations}
Specifically, \eqref{DOP-1b} contains the transmit power constraint and the power order constraint for SIC, where $P_a^{\max}$ is the maximum transmit power of Alice; \eqref{DOP-1c} guarantees that the SIC at Bob can be successfully carried out; \eqref{DOP-1d} denotes the QoS constraint with $R_{r,s_r(k)}^{\min}$ being the minimum rate requirement of Roy in downlink NOMA; \eqref{DOP-1e} is the covert constraint with $\varepsilon$ being predefined to specify a certain covertness; and \eqref{DOP-1f} describes the phase-shift constraints of the IRS elements. The optimization problem, given its original formulation in (P1), is challenging to solve, due to the highly coupled optimization variables $P_r$, $P_b$, and $\mathbf{\Theta}$ in the objective function and constraint \eqref{DOP-1d}, the non-convex constraint \eqref{DOP-1c}, and the complicated expression $\bar{\xi}_{w,\text{dl}}^\ast$ in \eqref{DOP-1e}. To tackle this challenge, in the following, we propose to optimize the transmit power and passive beamforming alternatively, until the convergence is achieved.


Due to the fact that $\beta(t)=\log\big(1+\frac{P_rt}{P_bt+\sigma_0^2}\big)$ is an increasing function with $t$, \eqref{DOP-1c} is equivalent to $|g_{ar}|^2\leq|g_{ab}|^2$, which is independent of the transmit power $P_r$ and $P_b$. Then, for any given phase shift $\mathbf{\Theta}$, problem (P1) is reduced to a transmit power optimization problem as
\begin{subequations}
  \begin{align}
    \text{(P2):}\ &\max_{P_r,P_b}\ R_{b,s_b(k)} \label{PA-1a}\\
    &\text{s.t.}\,\,  P_r+P_b\leq P_a^{\max},\ P_r\geq P_b, \label{PA-1b}\\
    &\quad\,\ P_r|g_{ar}|^2\geq\big(2^{R_{r,s_r(k)}^{\min}}-1\big) \big(P_b|g_{ar}|^2+\sigma_0^2\big), \label{PA-1c}\\
    &\quad\,\ \bar{\xi}_{w,\text{dl}}^\ast\geq1-\varepsilon,\ \varepsilon\in(0,1), \label{PA-1d}
  \end{align}
\end{subequations}
where $\gamma_\mathrm{th}=2^{R_{r,s_r(k)}^{\min}}-1$.
\begin{proposition}
  The maximum objective value of (P2) is achieved only when $P_r+P_b\leq P_a^{\max}$ in \eqref{PA-1b} holds with equality, i.e., $P_r+P_b=P_a^{\max}$.
\end{proposition}
\begin{proposition}
  With $P_r+P_b=P_a^{\max}$, $\bar{\xi}_{w,\text{dl}}^\ast$ is monotonically decreasing with respect to $P_b$.
\end{proposition}
\begin{IEEEproof}
  Please refer to Appendix C.
\end{IEEEproof}

With Propositions 1 and 2, problem (P2) can be simplified as
\begin{subequations}
  \begin{align}
    \text{(P3):}\ &\max_{P_b}\ R_{b,s_b(k)} \label{S-PA-1a}\\
    &\text{s.t.}\,\,  P_b\leq \min\Big\{\frac{P_a^{\max}}{2},\Xi,\bar{\xi}_{w,\text{dl}}^{\ast(-1)} (\varepsilon)\Big\},\ \varepsilon\in(0,1), \label{S-PA-1b}
  \end{align}
\end{subequations}
where $\Xi=\frac{P_a^{\max}|g_{ar}|^2-\gamma_\mathrm{th}\sigma_0^2} {(1+\gamma_\mathrm{th})|g_{ar}|^2}$. It is not difficult to verify that $R_{b,s_b(k)}$ is monotonically increasing with respect to $P_b$, and thus, the optimal power allocation coefficients to maximize $R_{b,s_b(k)}$ are obtained in closed-form as $P_b^\ast=\min\big\{\frac{P_a^{\max}}{2},\Xi, \bar{\xi}_{w,\text{dl}}^{\ast(-1)}(\varepsilon)\big\}$ and $P_r^\ast=P_a^{\max}-P_b^\ast$. In particular, $\bar{\xi}_{w,\text{dl}}^{\ast(-1)}(\varepsilon)$ can be computed numerically using MATLAB.


On the other hand, for any given feasible power allocation coefficients $P_r$ and $P_b$, problem (P1) is reduced to the following passive beamforming optimization problem
\begin{subequations}
  \begin{align}
    \text{(P4):}\ &\max_{\mathbf{\Theta}}\ R_{b,s_b(k)} \label{BF-1a}\\
    &\text{s.t.}\,\,  |g_{ar}|^2\leq|g_{ab}|^2, \label{BF-1b}\\
    &\quad\,\ \big(P_r-\gamma_\mathrm{th}P_b\big) |g_{ar}|^2\geq\gamma_\mathrm{th}\sigma_0^2, \label{BF-1c}\\
    &\quad\,\ 0\leq\theta_n\leq2\pi,\ \forall n. \label{BF-1d}
  \end{align}
\end{subequations}
Let $\mathbf{u}=[u_1,\dots,u_N]^H$ with $u_n=e^{j\theta_n}$, $\forall n$. Then, constraints in \eqref{BF-1d} are equivalent to $|u_n|^2=1$, $\forall n$. By applying the change of variables $\mathbf{\Lambda}_i=\frac{\mathrm{diag}(\mathbf{h}_a^H)\mathbf{h}_i}{\sqrt{L(d_a)L(d_i)}}$ and $v_i=\frac{h_{ai}}{\sqrt{L(d_{ai})}}$ with $i\in\{r,b\}$, as well as introducing
\begin{equation}
  \mathbf{H}_i=\bigg[\begin{array}{cc}
    \mathbf{\Lambda}_i\mathbf{\Lambda}_i^H & \mathbf{\Lambda}_iv_i\\
    v_i\mathbf{\Lambda}_i^H & 0
  \end{array}\bigg]\ \text{and}\ \bar{\mathbf{u}}=\bigg[\begin{array}{cc}
    \mathbf{u} \\
    1
  \end{array}\bigg],
\end{equation}
we obtain that $|g_{ai}|^2=|\mathbf{u}^H\mathbf{\Lambda}_i+v_i|^2 =\bar{\mathbf{u}}^H\mathbf{H}_i\bar{\mathbf{u}}+|v_i|^2 =\mathrm{tr}(\mathbf{H}_i\mathbf{U})+|v_i|^2$, where $\mathbf{U}=\bar{\mathbf{u}}\bar{\mathbf{u}}^H\succeq\mathbf{0}$ and $\mathrm{rank}(\mathbf{U})=1$.
With this result, problem (P4) can be equivalently transformed into
\begin{subequations}
  \begin{align}
    \text{(P5):}\ &\max_{\mathbf{U}}\ \mathrm{tr}(\mathbf{H}_b\mathbf{U}) \label{BF-2a}\\
    &\text{s.t.}\,\,  \mathrm{tr}(\mathbf{H}_r\mathbf{U})+|v_r|^2\leq \mathrm{tr}(\mathbf{H}_b\mathbf{U})+|v_b|^2, \label{BF-2b}\\
    &\quad\,\ \big(P_r-\gamma_\mathrm{th}P_b\big) \big(\mathrm{tr}(\mathbf{H}_r\mathbf{U})+|v_r|^2\big) \geq\gamma_\mathrm{th}\sigma_0^2, \label{BF-2c}\\
    &\quad\,\ \mathbf{U}_{n,n}=1,\ n=1,\dots,N+1, \label{BF-2d}\\
    &\quad\,\ \mathbf{U}\succeq\mathbf{0}, \label{BF-2e}\\
    &\quad\,\ \mathrm{rank}(\mathbf{U})=1. \label{BF-2f}
  \end{align}
\end{subequations}
Specifically, we use the fact in \eqref{BF-2a} that the function $\log\big(1+\frac{x}{\sigma_0^2}\big)$ is monotonically increasing with $x$, and thus, maximizing $\log\big(1+\frac{x}{\sigma_0^2}\big)$ is equivalent to maximizing $x$. However, problem (P5) is still non-convex due to the rank-one constraint \eqref{BF-2f}. To address this issue,
we apply SDR to relax the rank-one constraint \eqref{BF-2f}. Consequently, problem (P5) is reduced to
\begin{subequations}
  \begin{align}
    \text{(P6):}\ &\max_{\mathbf{U}}\ \mathrm{tr}(\mathbf{H}_b\mathbf{U}) \label{BF-3a}\\
    &\text{s.t.}\,\,  \eqref{BF-2b},\ \eqref{BF-2c},\ \eqref{BF-2d},\ \eqref{BF-2e}. \label{BF-3b}
  \end{align}
\end{subequations}
It is clear that problem (P6) is a convex semidefinite program (SDP), which can be efficiently solved using standard convex optimization solvers such as CVX \cite{CVX}. Generally, the relaxed problem (P6) may not always yield a rank-one solution, i.e., $\mathrm{rank}(\mathbf{U})\neq1$, which indicates that the optimal objective value of problem (P6) is an upper bound of problem (P5). Hence, we apply the Gaussian randomization method \cite{QingqingWu_TWC2019} to obtain an approximate rank-one solution from the higher-rank solution to problem (P6). The overall algorithm is given in Algorithm~\ref{alg:1} and its convergence performance is characterized in the following proposition.

\begin{proposition}
  The developed Algorithm~\ref{alg:1} is guaranteed to converge.
\end{proposition}
\begin{IEEEproof}
  Please refer to Appendix D.
\end{IEEEproof}

\begin{algorithm}[t]
  \caption{Alternating Optimization Algorithm}
  \label{alg:1}
  \begin{spacing}{1.1}
  \begin{algorithmic}[1]
    \STATE Initialize $\mathbf{\Theta}(1)$, $P_b(1)$, and $P_r(1)$. Set iteration index $t=1$.
    \REPEAT
    \STATE Given $\mathbf{\Theta}(t)$, solve problem (P3) to obtain $P_b(t+1)$ and $P_r(t+1)$.
    \STATE Given $P_b(t+1)$ and $P_r(t+1)$, solve problem (P6) to obtain $\mathbf{U}(t+1)$.
    \STATE Update $t=t+1$.
    \UNTIL{The increase of the objective value is below a threshold $\varrho>0$.}
    \IF{$\mathrm{rank}\big(\mathbf{U}(t+1)\big)=1$}
    \STATE Compute the eigenvalue decomposition of $\mathbf{U}(t+1)$ to obtain the nonzero eigenvalue $\lambda_\text{eigen}$ and the respective eigenvector $\mathbf{v}_\text{eigen}$.
    \RETURN $\mathbf{\Theta}(t+1)=\mathrm{diag}(\sqrt{\lambda_\text{eigen}} \mathbf{v}_\text{eigen})$.
    \ELSE
    \FOR{$q=1,\dots,Q$}
    \STATE Compute $\mathbf{U}(t+1)=\mathbf{V}\mathbf{\Sigma}\mathbf{V}^H$ and generate $\mathbf{e}_q=\mathbf{V}\mathbf{\Sigma}^{\frac12}\mathbf{r}_q$, where $\mathbf{r}_q\sim\mathcal{CN}(\mathbf{0}_{N+1},\mathbf{I}_{N+1})$.
    \STATE Compute $\mathbf{\Theta}_q=\mathrm{diag} \big(e^{j\arg(\frac{\mathbf{e}_q[1]}{\mathbf{e}_q[N+1]})},\dots, e^{j\arg(\frac{\mathbf{e}_q[N]}{\mathbf{e}_q[N+1]})}\big)$.
    \STATE Obtain the objective value of problem (P5), denoted by $R_q$.
    \ENDFOR
    \ENDIF
    \RETURN $\mathbf{\Theta}(t+1)=\arg\max_{q=1,\dots,Q}R_q$.
  \end{algorithmic}
  \end{spacing}
\end{algorithm}

\section{Covert Communication in IRS-Assisted Uplink NOMA}
\label{sec:uplink}

This section develops a new IRS-assisted uplink NOMA scheme to achieve covert wireless communication. From the worst-case perspective that Willie can optimally select his detection threshold, the minimum average detection probability of the proposed scheme is derived in closed form. A joint transmit power and passive beamforming optimization problem is formulated and solved by an effective alternating optimization algorithm.

\subsection{IRS-Assisted Uplink NOMA Scheme}

As illustrated in Fig.~\ref{fig:system-model}(b), using the uplink NOMA principle, Roy transmits his public signal $s_{\tilde r}(k)$ and Bob transmits his covert signal $s_{\tilde b}(k)$ to Alice simultaneously, where the public signal transmission of Roy is exploited as a cover for the covert signal transmission of Bob. Therefore, the received signal at Alice is given by
\begin{align}
\label{y-alice}
  y_{\tilde a}(k)&=\bigg(\frac{h_{ar}}{\sqrt{L(d_{ar})}} +\frac{\mathbf{h}_r^H\mathbf{\Phi}\mathbf{h}_a}{\sqrt{L(d_a) L(d_r)}}\bigg)\sqrt{P_{\tilde r}}s_{\tilde r}(k)\nonumber\\
  &\quad+\bigg(\frac{h_{ab}}{\sqrt{L(d_{ab})}}+ \frac{\mathbf{h}_b^H\mathbf{\Phi}\mathbf{h}_a}{\sqrt{L(d_a)L(d_b)}}\bigg) \sqrt{P_{\tilde b}}s_{\tilde b}(k)+n_a(k),
\end{align}
where $\mathbf{\Phi}=\text{diag}(e^{j\varphi_1},\dots,e^{j\varphi_N})\in\mathbb{C}^{N\times N}$ denotes the IRS diagonal reflection-coefficients matrix of uplink NOMA with $\varphi_n\in[0,2\pi)$, $P_{\tilde r}$ and $P_{\tilde b}$ denote the average transmit powers of Roy and Bob, and $n_a(k)$ denotes the AWGN at Alice. The optimal $\mathbf{\Phi}$ is designed using the CSI of the Roy/Bob-Alice and Roy/Bob-IRS-Alice links, which is detailed in Section IV-C. As indicated in \cite{ZengMing_CL2020}, for uplink NOMA, the users with better channel conditions are often decoded earlier. To facilitate the SIC decoding at Alice, without loss of generality, we assume that the users are arranged in a descending order according to their composite channels, i.e., $|g_{ra}|^2\geq|g_{ba}|^2$, where $g_{ra}=\frac{h_{ar}}{\sqrt{L(d_{ar})}} +\frac{\mathbf{h}_r^H\mathbf{\Phi}\mathbf{h}_a}{\sqrt{L(d_a)L(d_r)}}$ is the composite channel of Roy and $g_{ba}=\frac{h_{ab}}{\sqrt{L(d_{ab})}}+ \frac{\mathbf{h}_b^H\mathbf{\Phi}\mathbf{h}_a}{\sqrt{L(d_a)L(d_b)}}$ is the composite channel of Bob. With the above ordered channels, the covert signal is decoded at the last stage of SIC without suffering from severe inter-user interference, which is beneficial to Bob's covertness (i.e., the covert rate of Bob increases as the average transmit power increases). Then, the achievable rates of Alice to sequentially decode $s_{\tilde r}(k)$ and $s_{\tilde b}(k)$ are given by
\begin{align}
  R_{a,s_{\tilde r}(k)}&=\log\bigg(1+\frac{P_{\tilde r}|g_{ra}|^2}{P_{\tilde b}|g_{ba}|^2+\sigma_0^2}\bigg), \label{UL-rate-Alice-sr}\\
  R_{a,s_{\tilde b}(k)}&=\log\bigg(1+\frac{P_{\tilde b}|g_{ba}|^2}{\sigma_0^2}\bigg). \label{UL-rate-Alice-sb}
\end{align}

Similar to the downlink scenario, Willie performs the Neyman-Pearson test to judge whether Bob is transmitting a covert signal to Alice or not. As a result, the received signals under the null hypothesis $\mathcal{H}_0$ and the alternative hypothesis $\mathcal{H}_1$ are given, respectively, by
\begin{align}
  \mathcal{H}_0:\ y_{\tilde w}(k)&=\bigg(\frac{h_{rw}}{\sqrt{L(d_{rw})}}+ \frac{\mathbf{h}_r^H\mathbf{\Phi}\mathbf{h}_w}{\sqrt{L(d_r)L(d_w)}}\bigg) \sqrt{P_{\tilde r}}s_{\tilde r}(k)+z_w(k), \label{UL-yw0}\\
  \mathcal{H}_1:\ y_{\tilde w}(k)&=\bigg(\frac{h_{rw}}{\sqrt{L(d_{rw})}}+ \frac{\mathbf{h}_r^H\mathbf{\Phi}\mathbf{h}_w}{\sqrt{L(d_r)L(d_w)}}\bigg) \sqrt{P_{\tilde r}}s_{\tilde r}(k)\nonumber\\
  &\quad+\bigg(\frac{h_{bw}}{\sqrt{L(d_{bw})}}+ \frac{\mathbf{h}_b^H\mathbf{\Phi}\mathbf{h}_w}{\sqrt{L(d_b)L(d_w)}}\bigg)
  \sqrt{P_{\tilde b}}s_{\tilde b}(k)+z_w(k). \label{UL-yw1}
\end{align}
With the received signals in \eqref{UL-yw0} and \eqref{UL-yw1}, the average received power at Willie of the proposed IRS-assisted uplink NOMA scheme can be obtained as
\begin{equation}
  P_{\tilde w}=\begin{cases}
    \Big(\frac{|h_{rw}|^2}{L(d_{rw})}+\frac{|\mathbf{h}_r^H\mathbf{\Phi}\mathbf{h}_w|^2} {L(d_r)L(d_w)}\Big)P_{\tilde r}+\sigma_0^2, & \mathcal{H}_0,\\
    \Big(\frac{|h_{rw}|^2}{L(d_{rw})} +\frac{|\mathbf{h}_r^H\mathbf{\Phi}\mathbf{h}_w|^2}{L(d_r) L(d_w)}\Big)P_{\tilde r}+\Big(\frac{|h_{bw}|^2}{L(d_{bw})} +\frac{|\mathbf{h}_b^H\mathbf{\Phi}\mathbf{h}_w|^2}{L(d_b) L(d_w)}\Big)P_{\tilde b}+\sigma_0^2, & \mathcal{H}_1.
  \end{cases}
\end{equation}

\begin{remark}
  The optimal phase shift $\mathbf{\Phi}$ is selected based on the CSI of the Roy/Bob-Alice and Roy/Bob-IRS-Alice links, which helps create the IRS phase-shift uncertainty to Willie to achieve Bob's covert communication, e.g., both $|\mathbf{h}_r^H\mathbf{\Phi}\mathbf{h}_w|^2$ and $|\mathbf{h}_b^H\mathbf{\Phi}\mathbf{h}_w|^2$ serve as RVs at Willie.
\end{remark}

\subsection{Detection Error Probability of Willie}

Let $\zeta_{N,1}=\mathbf{h}_r^H\mathbf{\Phi}\mathbf{h}_w =\sum_{n=1}^N|h_{rn}||h_{wn}|e^{-j\chi_n}$ and $\zeta_{N,2}=\mathbf{h}_b^H\mathbf{\Phi}\mathbf{h}_w =\sum_{n=1}^N|h_{bn}||h_{wn}|e^{-j\chi_n'}$, where $h_{rn}$ and $h_{bn}$ are the fading coefficients from Roy and Bob to the $n$th element of the IRS, $\chi_n=\varphi_n^\ast+\arg(h_{rn})+\arg(h_{wn})$, and $\chi_n'=\varphi_n^\ast+\arg(h_{bn})+\arg(h_{wn})$. In particular, $\varphi_n^\ast$ is the optimal phase shift of the IRS in the uplink, which is a function of $\arg(h_{an})$, $\arg(h_{bn})$, $\arg(h_{rn})$, $\arg(h_{ab})$, and $\arg(h_{ar})$. The following lemma provides the statistics of $\chi_n$ and $\chi_n'$.
\begin{lemma}
  The phase values $\chi_n$ and $\chi_n'$ follow the same independent and uniform distribution on $[0,2\pi)$, namely, $f_{\chi_n}(x)=f_{\chi_n'}(x)=\frac1{2\pi}$ with $x\in[0,2\pi)$.
\end{lemma}
\begin{IEEEproof}
  The proof is similar to that in Appendix A, which is omitted for brevity.
\end{IEEEproof}

With Lemma 2, and again applying \cite[Lemma~2]{ZDing_WCL2020}, we can approximate the CDF of $\zeta_{N,1}$ and $\zeta_{N,2}$ as complex Gaussian RVs with zero mean and variance equal to $N$, i.e., $\zeta_{N,1},\zeta_{N,2}\sim\mathcal{CN}(0,N)$. Consequently, the false alarm probability of the IRS-assisted uplink NOMA scheme can be calculated as
\begin{align}
\label{FA-uplink}
  \mathbb{P}_\text{FA}'&=\Pr\bigg(\frac{P_{\tilde r}|h_{rw}|^2}{L(d_{rw})}+ \frac{P_{\tilde r}\zeta_{N,1}^2}{L(d_r)L(d_w)}+\sigma_0^2>\tau_\text{ul}\bigg) \nonumber\\
  &=\begin{cases}
    \underbrace{e^{\frac{P_{\tilde r}|h_{rw}|^2/L(d_{rw})+\sigma_0^2-\tau_\text{ul}} {P_{\tilde r}N/(L(d_r)L(d_w))}}}_{\nu_3}, &\text{if}\ \tau_\text{ul}>\sigma_0^2+\frac{P_{\tilde r}|h_{rw}|^2}{L(d_{rw})}, \\
    1, &\text{otherwise}.
  \end{cases}
\end{align}
The miss detection probability of the IRS-assisted uplink NOMA scheme is derived as
\begin{align}
\label{MD-uplink}
  \mathbb{P}_\text{MD}'&=\Pr\bigg(\frac{P_{\tilde r}|h_{rw}|^2}{L(d_{rw})} +\frac{P_{\tilde b}|h_{bw}|^2}{L(d_{bw})}+\frac{P_{\tilde r}\zeta_{N,1}^2}{L(d_r)L(d_w)}+\frac{P_{\tilde b}\zeta_{N,2}^2}{L(d_b)L(d_w)}+\sigma_0^2<\tau_\text{ul}\bigg)\nonumber\\
  &=\begin{cases}
    1-\underbrace{e^{\frac{P_{\tilde r}|h_{rw}|^2/L(d_{rw})+P_{\tilde b}|h_{bw}|^2/L(d_{bw})+\sigma_0^2-\tau_\text{ul}} {(P_{\tilde r}/L(d_r)+P_{\tilde b}/L(d_b))N/L(d_w)}}}_{\nu_4}, & \text{if}\ \tau_\text{ul}>\sigma_0^2+\frac{P_{\tilde r}|h_{rw}|^2}{L(d_{rw})} +\frac{P_{\tilde b}|h_{bw}|^2}{L(d_{bw})},\\
    0, & \text{otherwise}.
  \end{cases}
\end{align}
Summarizing \eqref{FA-uplink} and \eqref{MD-uplink}, the detection error probability of the IRS-assisted uplink NOMA scheme is obtained as
\begin{equation}
\label{DEP-uplink}
  \xi_{w,\text{ul}}=\begin{cases}
    1, & \text{if}\ \tau_\text{ul}<\sigma_0^2+\frac{P_{\tilde r}|h_{rw}|^2}{L(d_{rw})},\\
    \nu_3, & \text{if}\ \sigma_0^2+\frac{P_{\tilde r}|h_{rw}|^2}{L(d_{rw})}\leq\tau_\text{ul}\leq \sigma_0^2+\frac{P_{\tilde r}|h_{rw}|^2}{L(d_{rw})}+\frac{P_{\tilde b}|h_{bw}|^2}{L(d_{bw})},\\
    1+\nu_3-\nu_4, & \text{if}\ \tau_\text{ul}>\sigma_0^2+\frac{P_{\tilde r}|h_{rw}|^2}{L(d_{rw})}+\frac{P_{\tilde b}|h_{bw}|^2}{L(d_{bw})}.
  \end{cases}
\end{equation}
The following theorem gives the optimal detection threshold $\tau_\text{ul}$ to minimize the detection error probability \eqref{DEP-uplink}.
\begin{theorem}
  The optimal detection threshold of Willie to minimize the detection error probability of the IRS-assisted uplink NOMA scheme is obtained as
  \begin{equation}
  \label{uplink-optimal-threshold}
    \tau_\text{ul}^\ast=\begin{cases}
      \sigma_0^2+\frac{P_{\tilde r}|h_{rw}|^2}{L(d_{rw})}+\frac{P_{\tilde b}|h_{bw}|^2}{L(d_{bw})}, & \text{if}\ |h_{bw}|^2\geq \frac{P_{\tilde r}N}{P_{\tilde b}\phi_3}\ln\big(\frac{P_{\tilde r}\phi_2+P_{\tilde b}}{P_{\tilde r}\phi_2}\big),\\
      \frac{P_{\tilde r}|h_{rw}|^2}{L(d_{rw})}-\frac{P_{\tilde r}\phi_2|h_{bw}|^2}{L(d_{bw})}+\sigma_0^2\\
      \quad+\frac{P_{\tilde r}(P_{\tilde r}\phi_2+P_{\tilde b})N}{P_{\tilde b}L(d_r)L(d_w)}\ln\big(\frac{P_{\tilde r}\phi_2+P_{\tilde b}}{P_{\tilde r}\phi_2}\big), & \text{otherwise},
    \end{cases}
  \end{equation}
  where $\phi_2=\frac{L(d_b)}{L(d_r)}$ and $\phi_3=\frac{L(d_r)L(d_w)}{L(d_{bw})}$.
\end{theorem}
\begin{IEEEproof}
  The proof is similar to that in Appendix B and thus omitted for simplicity.
\end{IEEEproof}

Substituting \eqref{uplink-optimal-threshold} into \eqref{DEP-uplink}, the minimum detection error probability of the IRS-assisted uplink NOMA scheme can be derived as
\begin{equation}
\label{minDEP-uplink}
  \xi_{w,\text{ul}}^\ast=\begin{cases}
    e^{-\frac{P_{\tilde b}\phi_3|h_{bw}|^2}{P_{\tilde r}N}}, & \text{if}\ |h_{bw}|^2\geq \frac{P_{\tilde r}N}{P_{\tilde b}\phi_3}\ln\big(\frac{P_{\tilde r}\phi_2+P_{\tilde b}}{P_{\tilde r}\phi_2}\big),\\
    1-\frac{P_{\tilde b}}{P_{\tilde r}\phi_2}\big(\frac{P_{\tilde r}\phi_2+P_{\tilde b}}{P_{\tilde r}\phi_2}\big)^{-\frac{P_{\tilde r}\phi_2+P_{\tilde b}}{P_{\tilde b}}} e^{\frac{\phi_4|h_{bw}|^2}{N}}, & \text{otherwise},
  \end{cases}
\end{equation}
where $\phi_4=\frac{L(d_b)L(d_w)}{L(d_{bw})}$. Based on this result, the minimum average detection error probability in closed-form is provided in the following theorem.

\begin{theorem}
  The minimum average detection error probability of Willie achieved by the IRS-assisted uplink NOMA scheme is expressed as
  \begin{align}
  \label{minDEP-uplink-closed}
    \bar{\xi}_{w,\text{ul}}^\ast&=\frac{\big(\frac{P_{\tilde r}\phi_2+P_{\tilde b}}{P_{\tilde r}\phi_2}\big)^ {-(1+\frac{P_{\tilde r}N}{P_{\tilde b}\phi_3})}}{P_{\tilde b}\phi_3(P_{\tilde r}N)^{-1}+1}+ 1-\Big(\frac{P_{\tilde r}\phi_2+P_{\tilde b}}{P_{\tilde r}\phi_2}\Big)^{-\frac{P_{\tilde r}N}{P_{\tilde b}\phi_3}} \nonumber\\
    &\quad-\frac{\frac{P_{\tilde b}}{P_{\tilde r}\phi_2}\big(\frac{P_{\tilde r}\phi_2+P_{\tilde b}}{P_{\tilde r}\phi_2}\big)^{-\frac{P_{\tilde r}\phi_2+P_{\tilde b}}{P_{\tilde b}}}}{\phi_4N^{-1}-1} \bigg[\Big(\frac{P_{\tilde r}\phi_2+P_{\tilde b}}{P_{\tilde r}\phi_2}\Big)^{\frac{P_{\tilde r}}{P_{\tilde b}}(\phi_2-\frac{N}{\phi_3})}-1\bigg].
  \end{align}
\end{theorem}
\begin{IEEEproof}
  The proof is similar to that of Theorem 2, which is omitted for brevity.
\end{IEEEproof}

\begin{remark}
  According to \eqref{minDEP-uplink-closed}, we have the following observations: 1) $\bar{\xi}_{w,\text{ul}}^\ast\rightarrow0$, if $P_{\tilde r}$ is finite but $P_{\tilde b}\rightarrow\infty$; and 2) $\bar{\xi}_{w,\text{ul}}^\ast\rightarrow1$, if $P_{\tilde b}$ is finite but $P_{\tilde r}\rightarrow\infty$.
\end{remark}

\subsection{Joint Power and Beamforming Optimization}

Similar to the downlink scenario, our goal is to maximize the covert rate of Bob via joint transmit power and passive beamforming optimization, subject to the individual transmit power constraints at Roy and Bob, the QoS constraint at Roy, and the minimum average detection error probability constraint at Willie. Accordingly, the optimization problem is formulated as
\begin{subequations}
  \begin{align}
    \text{(P7):} &\max_{P_{\tilde r},P_{\tilde b},\mathbf{\Phi}}\ R_{a,s_{\tilde b}(k)} \label{UOP-1a}\\
    &\text{s.t.}\,\,  P_{\tilde r}\leq P_r^{\max},\ P_{\tilde b}\leq P_b^{\max}, \label{UOP-1b}\\
    &\quad\ |g_{ra}|^2\geq|g_{ba}|^2, \label{UOP-1c}\\
    &\quad\ R_{a,s_{\tilde r}(k)}\geq R_{a,s_{\tilde r}(k)}^{\min}, \label{UOP-1d}\\
    &\quad\ \bar{\xi}_{w,\text{ul}}^\ast\geq1-\varepsilon,\ \varepsilon\in(0,1), \label{UOP-1e}\\
    &\quad\ 0\leq\varphi_n\leq2\pi,\ n=1,\dots,N. \label{UOP-1f}
  \end{align}
\end{subequations}
In \eqref{UOP-1b}, $P_r^{\max}$ and $P_b^{\max}$ denote the maximum transmit powers of Roy and Bob, respectively. In \eqref{UOP-1d}, $R_{a,s_{\tilde r}(k)}^{\min}$ denotes the minimum rate requirement of Roy in uplink NOMA. To handle the highly coupled optimization variables $P_{\tilde r}$, $P_{\tilde b}$, and $\mathbf{\Phi}$ in the objective function, we decompose problem (P7) into a transmit power optimization problem and a passive beamforming optimization problem, and apply the alternating optimization similar to that of the downlink scenario.


By fixing $\mathbf{\Phi}$, problem (P7) is reduced to a transmit power optimization problem as
\begin{subequations}
  \begin{align}
    \text{(P8):}\ &\max_{P_{\tilde r},P_{\tilde b}}\ R_{a,s_{\tilde b}(k)} \label{UL-PA-1a}\\
    &\text{s.t.}\,\,  P_{\tilde r}\leq P_r^{\max},\ P_{\tilde b}\leq P_b^{\max}, \label{UL-PA-1b}\\
    &\quad\,\ P_{\tilde r}|g_{ra}|^2\geq \tilde{\gamma}_\mathrm{th}P_{\tilde b}|g_{ba}|^2+\tilde{\gamma}_\mathrm{th}\sigma_0^2, \label{UL-PA-1c}\\
    &\quad\,\ \bar{\xi}_{w,\text{ul}}^\ast\geq1-\varepsilon,\ \varepsilon\in(0,1), \label{UL-PA-1d}
  \end{align}
\end{subequations}
where $\tilde{\gamma}_\mathrm{th}=2^{R_{a,s_{\tilde r}(k)}^{\min}}-1$.
 The optimal solution to problem (P8) is provided in the following proposition.
\begin{proposition}
  The optimal transmit power allocation coefficients to problem (P8) are given by $P_{\tilde r}^\ast=P_r^{\max}$ and $P_{\tilde b}^\ast=\min\big\{\bar{\xi}_{w,\text{ul}}^{\ast(-1)}(\varepsilon), \frac{P_r^{\max}|g_{ra}|^2-\tilde{\gamma}_\mathrm{th}\sigma_0^2} {\tilde{\gamma}_\mathrm{th}|g_{ba}|^2},P_b^{\max}\big\}$.
\end{proposition}
\begin{IEEEproof}
  Similar to that in Appendix C.
\end{IEEEproof}

On the other hand, for given $P_{\tilde r}$ and $P_{\tilde b}$, problem (P7) is reduced to a passive beamforming optimization problem as
\begin{subequations}
  \begin{align}
    \text{(P9):}\ &\max_{\mathbf{\Phi}}\ R_{a,s_{\tilde b}(k)} \label{UL-BF-1a}\\
    &\text{s.t.}\,\, |g_{ra}|^2\geq|g_{ba}|^2, \label{UL-BF-1b}\\
    &\quad\,\ P_{\tilde r}|g_{ra}|^2\geq \tilde{\gamma}_\mathrm{th}P_{\tilde b}|g_{ba}|^2+\tilde{\gamma}_\mathrm{th}\sigma_0^2, \label{UL-BF-1c}\\
    &\quad\,\ 0\leq\theta_n\leq2\pi,\ \forall n. \label{UL-BF-1d}
  \end{align}
\end{subequations}
Let $\mathbf{w}=[w_1,\dots,w_N]^H$ with $w_n=e^{j\varphi_n}$, constraint \eqref{UL-BF-1d} is transformed into $|w_n|^2=1$, $\forall n$. Then, we apply the change of variable $\tilde{\mathbf{\Lambda}}_i =\frac{\mathrm{diag}(\mathbf{h}_i^H)\mathbf{h}_a}{\sqrt{L(d_a)L(d_i)}}$, $i\in\{r,b\}$ and define the following auxiliary matrices:
\begin{equation}
  \tilde{\mathbf{H}}_i=\bigg[\begin{array}{cc}
    \tilde{\mathbf{\Lambda}}_i\tilde{\mathbf{\Lambda}}_i^H & \tilde{\mathbf{\Lambda}}_iv_i\\
    v_i\tilde{\mathbf{\Lambda}}_i^H & 0
  \end{array}\bigg]\ \text{and}\ \bar{\mathbf{w}}=\bigg[\begin{array}{cc}
    \mathbf{w} \\
    1
  \end{array}\bigg].
\end{equation}
We have $|g_{ia}|^2=|\mathbf{w}^H\tilde{\mathbf{\Lambda}}_i+v_i|^2 =\bar{\mathbf{w}}^H\tilde{\mathbf{H}}_i\bar{\mathbf{w}}+|v_i|^2$. As a result, problem (P9) can be equivalently formulated as
\begin{subequations}
  \begin{align}
    \text{(P10):}\ &\max_{\mathbf{W}}\ \mathrm{tr}(\tilde{\mathbf{H}}_b\mathbf{W}) \label{UL-BF-2a}\\
    &\text{s.t.}\,\, \mathrm{tr}(\tilde{\mathbf{H}}_r\mathbf{W})+|v_r|^2\geq \mathrm{tr}(\tilde{\mathbf{H}}_b\mathbf{W})+|v_b|^2, \label{UL-BF-2b}\\
    &\quad\,\ P_{\tilde r}\big(\mathrm{tr}(\tilde{\mathbf{H}}_r\mathbf{W})+|v_r|^2\big)\geq \tilde{\gamma}_\mathrm{th}P_{\tilde b}\big(\mathrm{tr}(\tilde{\mathbf{H}}_b\mathbf{W})+|v_b|^2\big) +\tilde{\gamma}_\mathrm{th}\sigma_0^2, \label{UL-BF-2c}\\
    &\quad\,\  \mathbf{W}_{n,n}=1,\ n=1,\dots,N+1, \label{UL-BF-2d}\\
    &\quad\,\ \mathbf{W}=\bar{\mathbf{w}}\bar{\mathbf{w}}^H\succeq\mathbf{0}, \label{UL-BF-2e}\\
    &\quad\,\ \mathrm{rank}(\mathbf{W})=1. \label{UL-BF-2f}
  \end{align}
\end{subequations}
By applying SDR to relax the rank-one constraint \eqref{UL-BF-2f}, problem (P10) is reduced to a convex SDP, which can be efficiently solved by CVX. In addition, the Gaussian randomization is used to extract an approximate rank-one solution. The overall algorithm for solving problem (P7) is similar to Algorithm~\ref{alg:1}, which thus is omitted.

\section{Numerical Studies}
\label{sec:simulation}

This section presents numerical results to evaluate the system performance achieved by the proposed IRS-assisted downlink and uplink NOMA schemes. For illustration, we assume that Alice, Bob, Roy, Willie, and the IRS (which forms a uniform rectangular array) are located at $(0,0)$, $(100,0)$, $(100,5)$, $(90,-5)$, and $(90,5)$ in meter (m) in a two-dimensional plane, respectively. The effective path loss function is modeled as $L(d)\ (\mathrm{in\ dB})=35.1+36.7\lg(d)-G_t-G_r$, where $G_t$ and $G_r$ denote the transmitter and receiver antenna gains with $G_t=G_r=10$ dBi. The variance of the AWGN is $\sigma_0^2=-80$ dBm. Moreover, the stopping threshold in Algorithm~\ref{alg:1} is $\varrho=10^{-4}$, and the Monte-Carlo simulations are averaged over $10^3$ independent trials.

\subsection{IRS-Assisted Downlink NOMA Scheme}

\begin{figure*}[t]
  \normalsize
  \centering
  \begin{minipage}[t]{0.48\textwidth}
  \centering
    \includegraphics[width=3.0in]{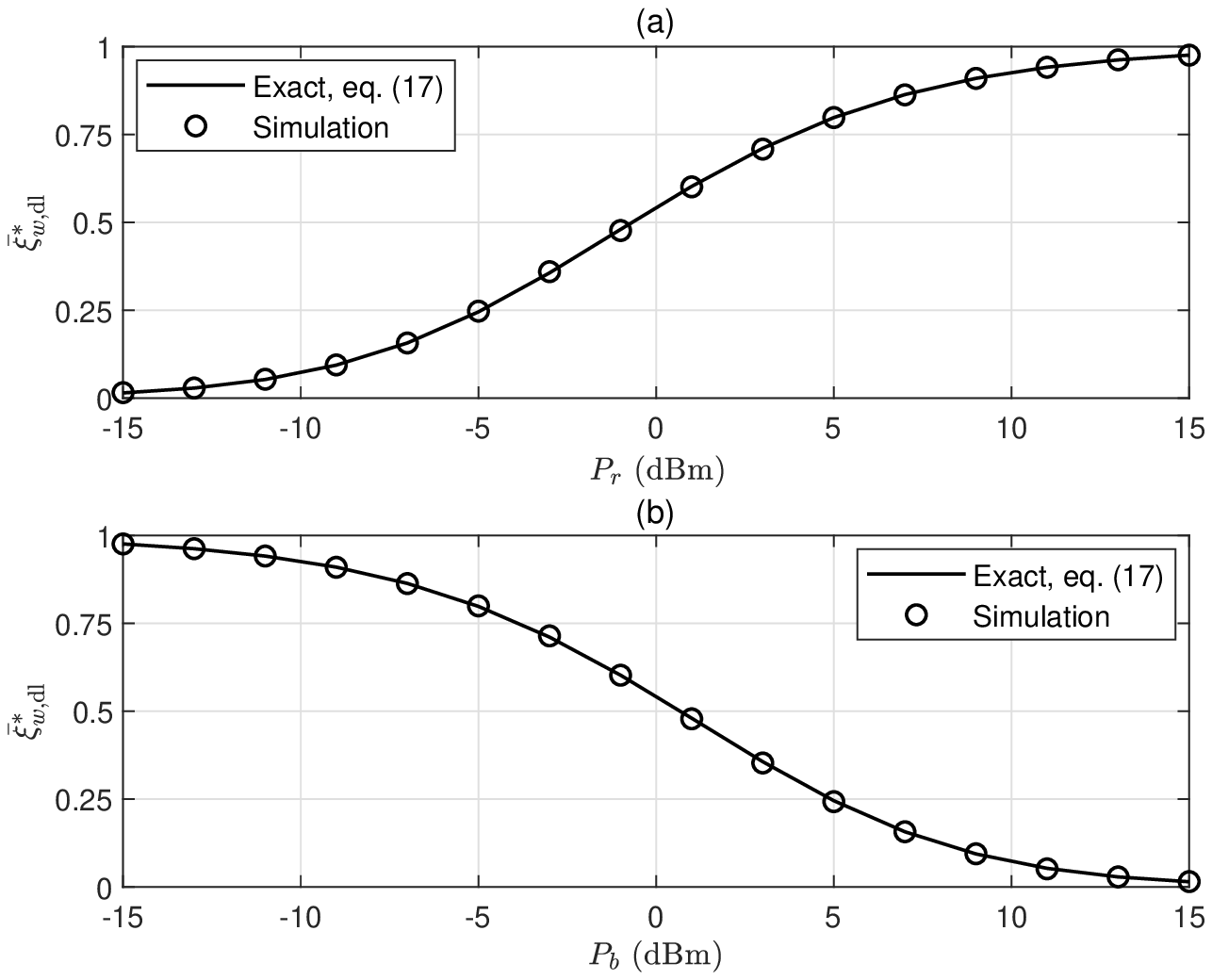}\vspace{-2mm}
   \caption{Minimum average detection probability versus the average transmit power in downlink, where $N=32$.}
   \label{fig:sim-1}\vspace{-3mm}
  \end{minipage}
  \hspace{1mm}
  \begin{minipage}[t]{0.48\textwidth}
  \centering
    \includegraphics[width=3.0in]{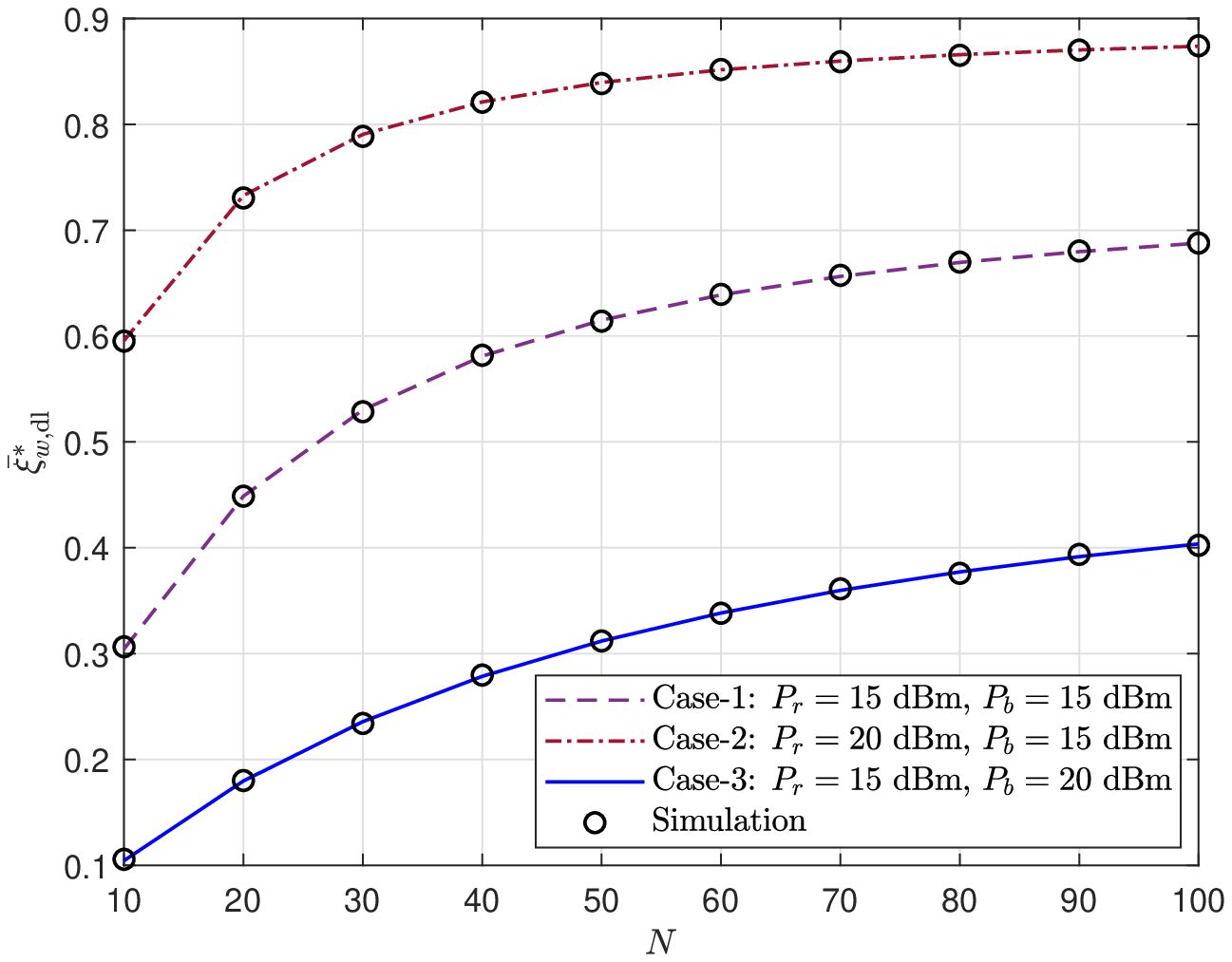}\vspace{-2mm}
    \caption{Minimum average detection probability versus the number of reflecting elements at the IRS in downlink.}
    \label{fig:sim-2}\vspace{-3mm}
  \end{minipage}
\end{figure*}

Fig.~\ref{fig:sim-1} shows the downlink minimum average detection probability $\bar{\xi}_{w,\text{dl}}^\ast$ as a function of the average transmit power $P_r$ for Roy's public signal in Fig.~\ref{fig:sim-1}(a), and the average transmit power $P_b$ for Bob's covert signal in Fig.~\ref{fig:sim-1}(b), respectively. It is found that $\bar{\xi}_{w,\text{dl}}^\ast$ monotonically increases with $P_r$, as seen in Fig.~\ref{fig:sim-1}(a), but monotonically decreases with $P_b$, as seen in Fig.~\ref{fig:sim-1}(b), which is proved in Proposition 2. The fundamental reason is that, in the proposed scheme, Bob exploits the public transmission of Roy as a cover for its covert communication. An increase in $P_r$ enhances the strength of the cover, and in turn creates a larger uncertainty to Willie, as shown in \eqref{Pw}. However, when $P_b$ increases, it becomes easier for Willie to detect Bob's covert communication. It is also observed from the figures that by further increasing $P_r$ and $P_b$, $\bar{\xi}_{w,\text{dl}}^\ast$ gradually goes to 1 and 0, respectively, which is consistent with our findings in Remark~3. Furthermore, in both Figs.~\ref{fig:sim-1}(a) and (b), the exact results in \eqref{minDEP-downlink-closed} match well with the simulated ones, thus verifying the accuracy of the derived analytical results.

In Fig.~\ref{fig:sim-2}, we plot the downlink minimum average detection probability $\bar{\xi}_{w,\text{dl}}^\ast$ as a function of the number of reflecting elements $N$ at the IRS. A general trend from the figure is that $\bar{\xi}_{w,\text{dl}}^\ast$ increases by increasing $N$, which is due to the fact that an increase in $N$ is helpful in creating a large IRS phase-shift uncertainty to confuse Willie. It is also observed from Fig.~\ref{fig:sim-2} that by fixing $P_b$, increasing $P_r$ helps in achieving a larger $\bar{\xi}_{w,\text{dl}}^\ast$, i.e., from Case-1 to Case-2. Under the same total transmit power, a smaller $P_b$ yields a larger $\bar{\xi}_{w,\text{dl}}^\ast$, i.e., from Case-3 to Case-2.

\begin{figure*}[t]
  \normalsize
  \centering
  \begin{minipage}[t]{0.48\textwidth}
  \centering
    \includegraphics[width=3.0in]{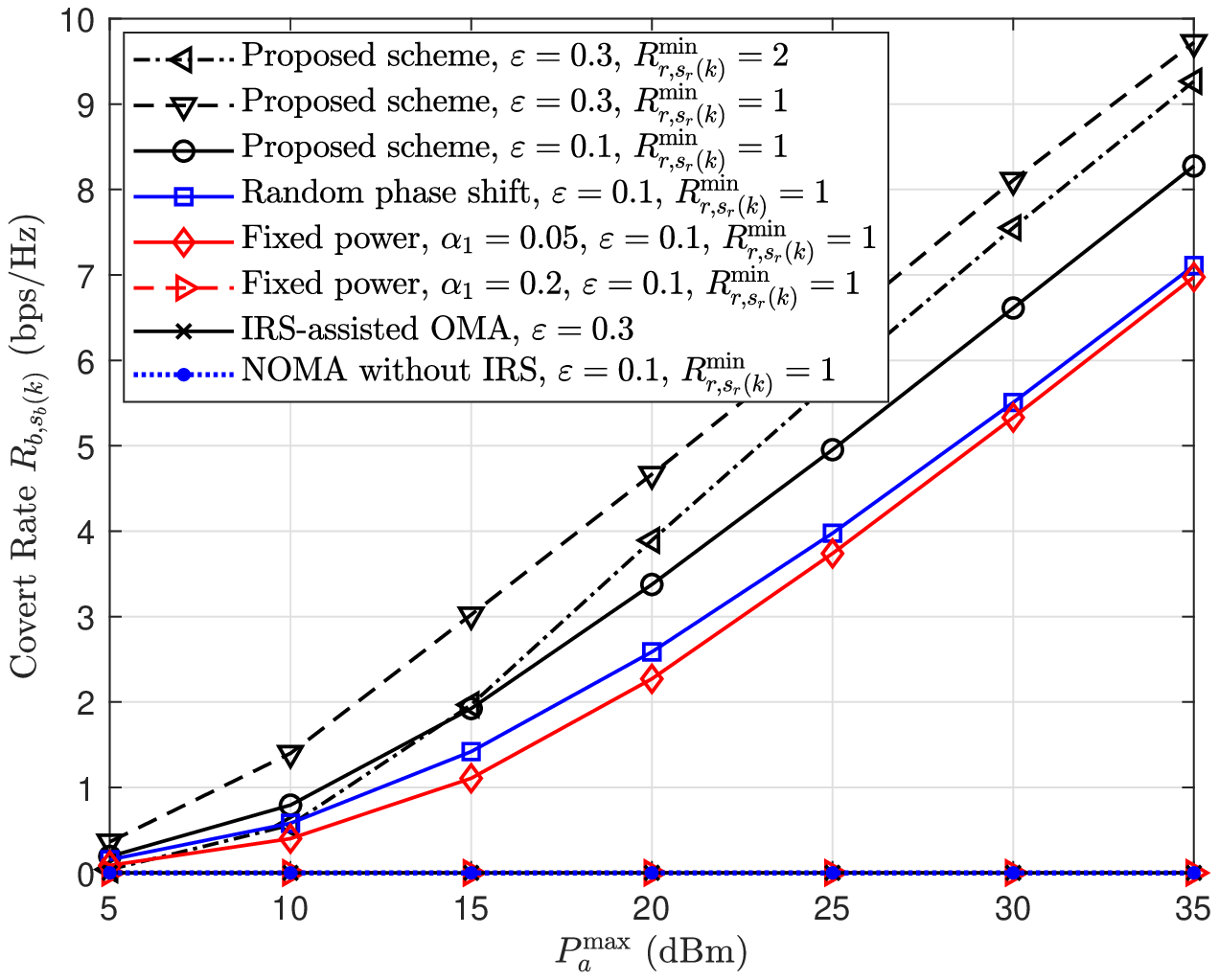}\vspace{-2mm}
   \caption{Covert rate versus the transmit power budget in downlink, where $N=32$.}
   \label{fig:sim-3}\vspace{-3mm}
  \end{minipage}
  \hspace{1mm}
  \begin{minipage}[t]{0.48\textwidth}
  \centering
    \includegraphics[width=3.0in]{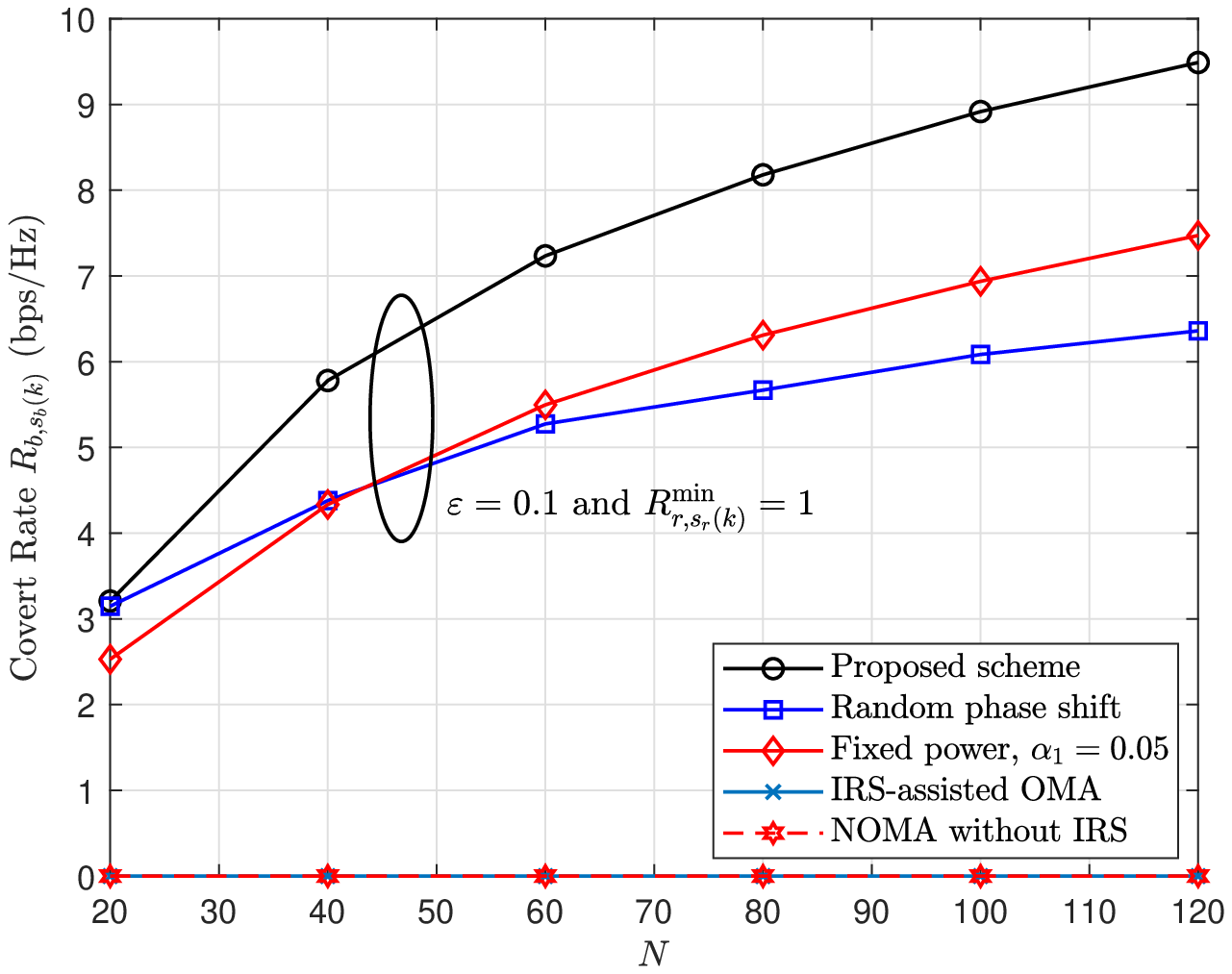}\vspace{-2mm}
    \caption{Covert rate versus the number of reflecting elements at the IRS in downlink, where $P_a^{\max}=25$ dBm.}
    \label{fig:sim-4}\vspace{-3mm}
  \end{minipage}
\end{figure*}

Figs.~\ref{fig:sim-3} and \ref{fig:sim-4} plot the covert rate of Bob achieved by different schemes versus the transmit power budget at Alice. We compare the following benchmark schemes: 1) Random phase shift: we randomly select the phase-shift values of the IRS in $[0,2\pi)$ and perform the optimal power allocation at Alice; 2) Fixed power: we set the transmit power allocation at Alice as $P_b=\alpha_1P_a^{\max}$ and $P_r=(1-\alpha_1)P_a^{\max}$ with optimized phase shifts of the IRS; 3) IRS-assisted OMA; 4) NOMA without IRS. From both figures, it is observed that the proposed scheme significantly outperforms the four other benchmark schemes in terms of achieving a higher covert rate. In Fig.~\ref{fig:sim-3}, for the fixed power with an incorrect choice of $P_b$ (e.g., when $\alpha_1=0.2$), the covert rate will always be zero, since the QoS constraint at Roy or the minimum average detection error probability constraint at Willie is not satisfied in this case. In Fig.~\ref{fig:sim-4}, the covert rate of the random phase shift increases much more slowly than that of the proposed scheme as $N$ increases, due to the passive beamforming gain loss with random phase shifts. All these observations demonstrate the effectiveness of the joint power and beamforming design, that can dynamically adjust Alice's transmit power and the IRS's phase shifts to meet the QoS requirement and the covertness constraint, to maximize the covert rate at Bob. It can be also seen from Fig.~\ref{fig:sim-3} that the covert rate of the proposed scheme decreases as $\varepsilon$ decreases or $R_{r,s_r(k)}^{\min}$ increases. The main reason is that, the covertness constraint at Willie and the QoS constraint at Roy become more stringent with the decreased $\varepsilon$ and the increased $R_{r,s_r(k)}^{\min}$, which results in a smaller covert rate for Bob. Furthermore, the IRS-assisted OMA and NOMA without IRS schemes always achieve the zero covert rate, which confirms our findings in Remark 2.


\subsection{IRS-Assisted Uplink NOMA Scheme}

\begin{figure*}[t]
  \normalsize
  \centering
  \begin{minipage}[t]{0.48\textwidth}
  \centering
    \includegraphics[width=3.0in]{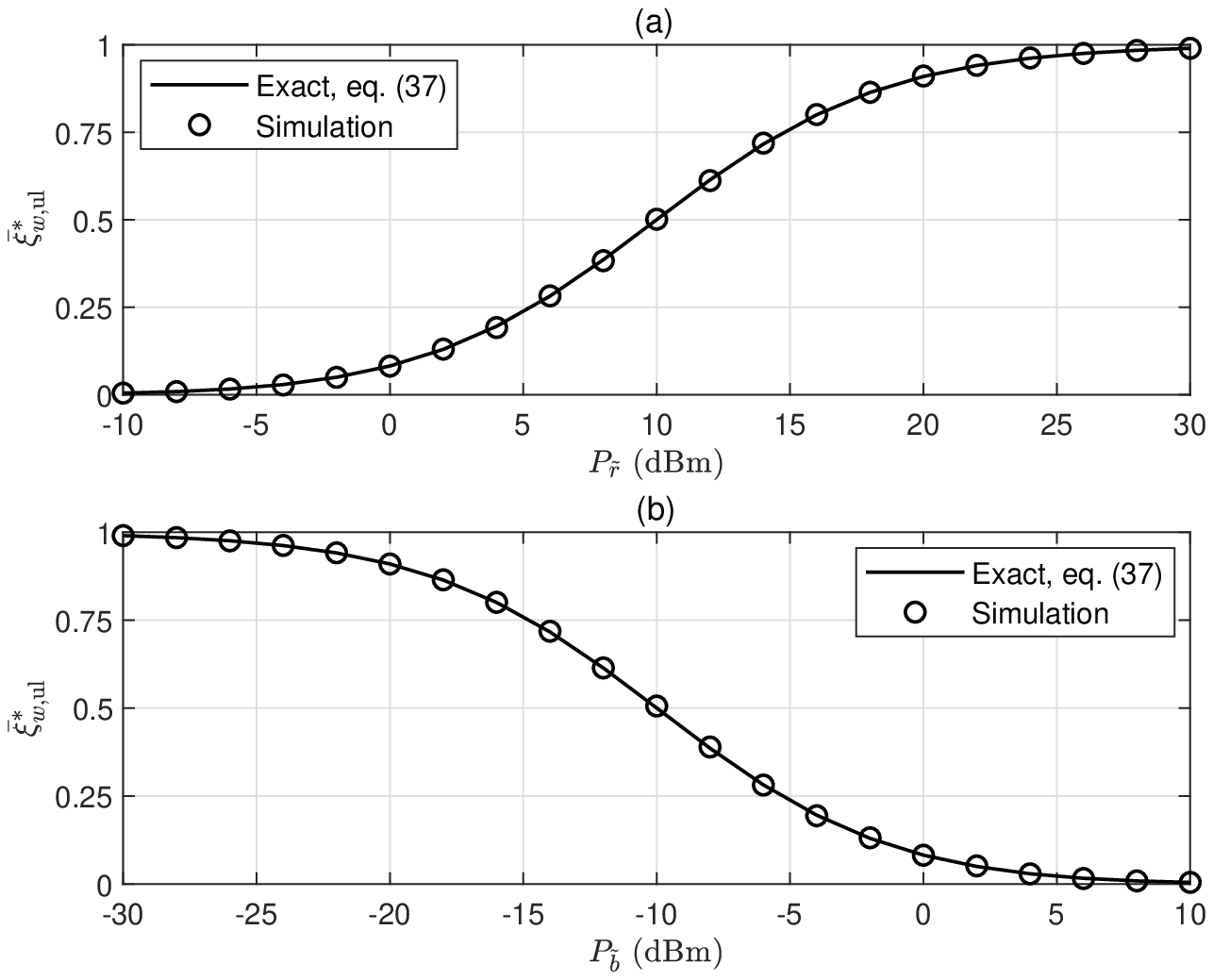}\vspace{-2mm}
   \caption{Minimum average detection probability versus the average transmit power in uplink, where $N=32$.}
   \label{fig:sim-5}\vspace{-3mm}
  \end{minipage}
  \hspace{1mm}
  \begin{minipage}[t]{0.48\textwidth}
  \centering
    \includegraphics[width=3.0in]{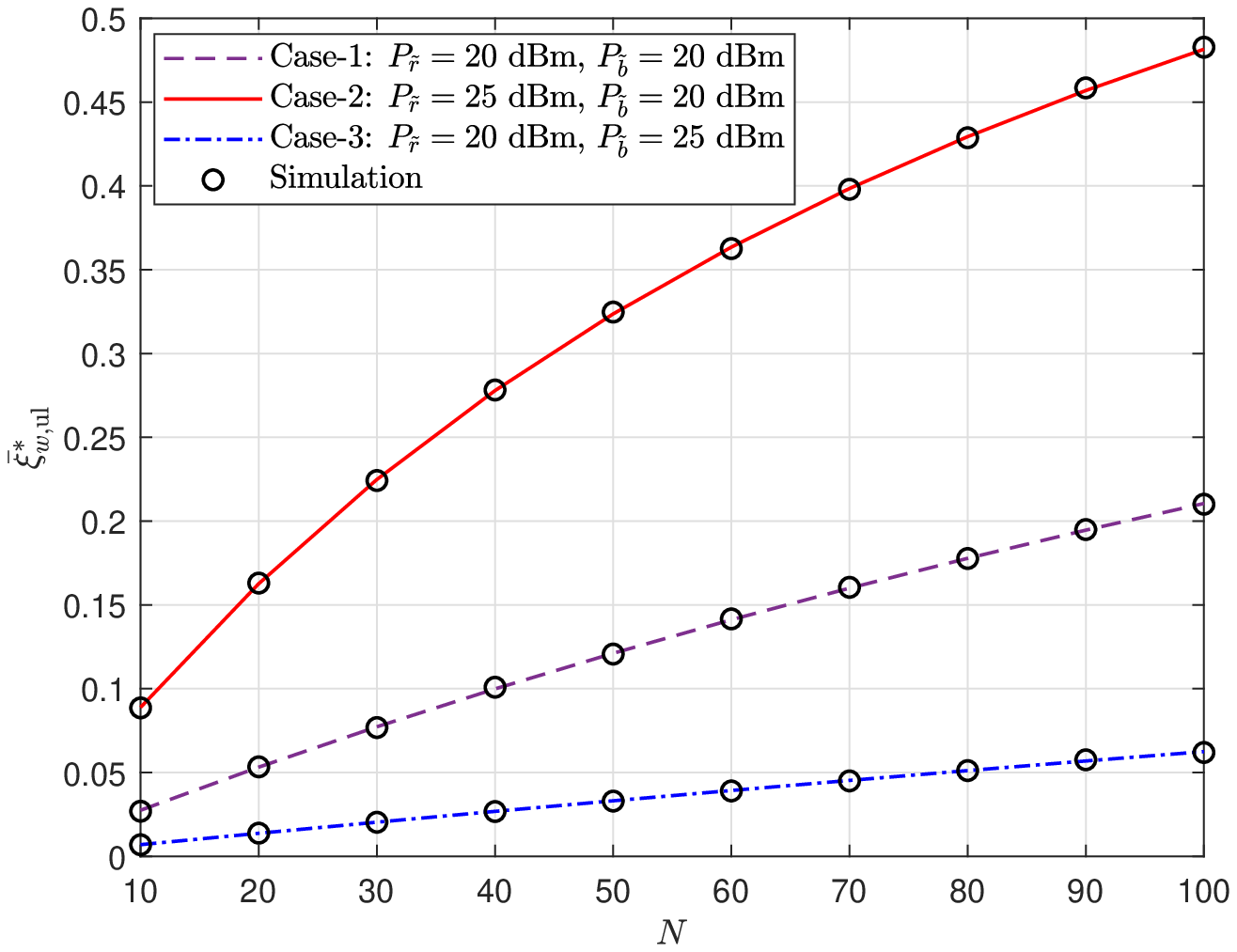}\vspace{-2mm}
    \caption{Minimum average detection probability versus the number of reflecting elements at the IRS in uplink.}
    \label{fig:sim-6}\vspace{-3mm}
  \end{minipage}
\end{figure*}

In Fig.~\ref{fig:sim-5}, the uplink minimum average detection probability $\bar{\xi}_{w,\text{ul}}^\ast$ versus the average transmit power of Roy $P_{\tilde r}$ and Bob $P_{\tilde b}$. From Fig.~\ref{fig:sim-5}(a) and (b), it is seen that $\bar{\xi}_{w,\text{ul}}^\ast$ monotonically increases with $P_{\tilde r}$ while decreasing with $P_{\tilde b}$, for which the reason is similar to that of the downlink NOMA scenario. Furthermore, by increasing $P_{\tilde r}$ and $P_{\tilde b}$, $\bar{\xi}_{w,\text{ul}}^\ast$ approaches 1 and 0, respectively, which confirms our results in Remark~5. In addition, the accuracy of the derived analytical results in Theorem~4 is also evaluated in these two subfigures.

Fig.~\ref{fig:sim-6} shows the uplink minimum average detection probability $\bar{\xi}_{w,\text{ul}}^\ast$ versus the number of reflecting elements $N$ at the IRS. As expected, $\bar{\xi}_{w,\text{ul}}^\ast$ increases with $N$ due to the significant phase-shift uncertainty of the IRS with a large $N$. It is also seen from this figure that a higher $P_{\tilde r}$ or a lower $P_{\tilde b}$ is beneficial to confound Willie, which yields a larger $\bar{\xi}_{w,\text{ul}}^\ast$.

\begin{figure*}[t]
  \normalsize
  \centering
  \begin{minipage}[t]{0.48\textwidth}
  \centering
    \includegraphics[width=3.0in]{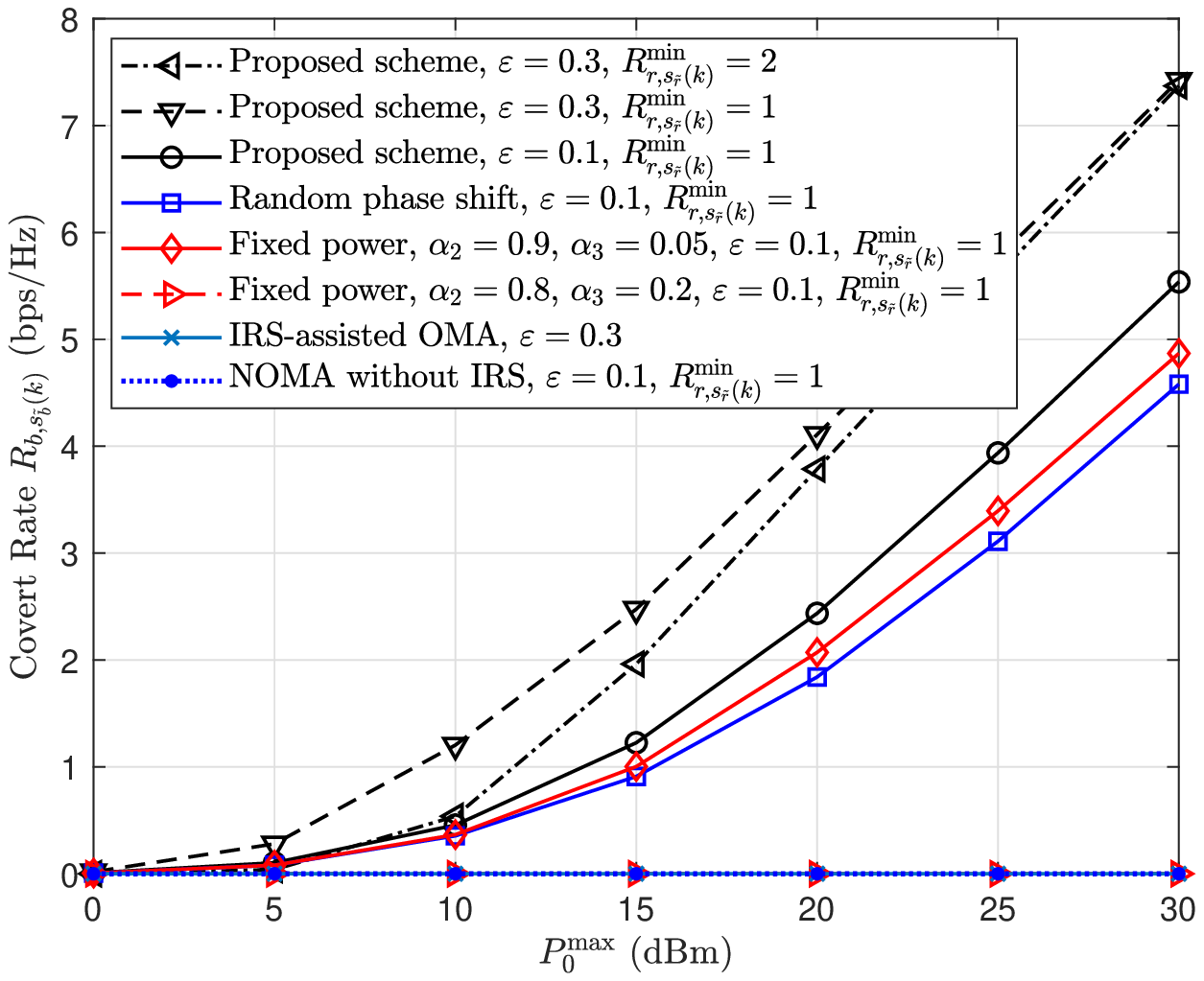}\vspace{-2mm}
   \caption{Covert rate versus the transmit power budget in uplink, where $N=32$ and $P_0^{\max}=P_r^{\max}=P_b^{\max}$.}
   \label{fig:sim-7}\vspace{-3mm}
  \end{minipage}
  \hspace{1mm}
  \begin{minipage}[t]{0.48\textwidth}
  \centering
    \includegraphics[width=3.0in]{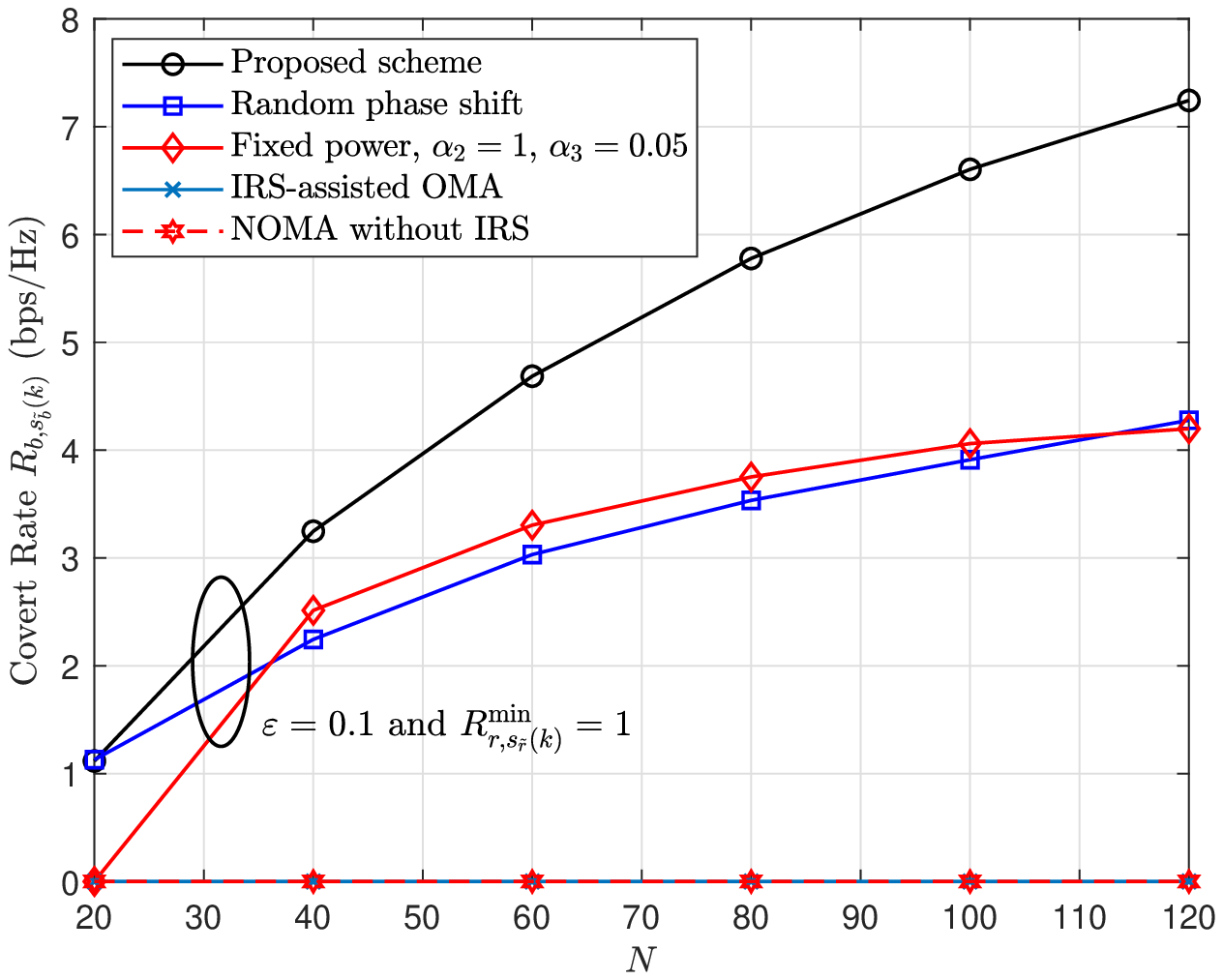}\vspace{-2mm}
    \caption{Covert rate versus the number of reflecting elements at the IRS in uplink, where $P_r^{\max}=P_b^{\max}=20$ dBm.}
    \label{fig:sim-8}\vspace{-3mm}
  \end{minipage}
\end{figure*}

Figs.~\ref{fig:sim-7} and \ref{fig:sim-8} show the uplink covert rate of Bob achieved by different schemes versus the common transmit power budget $P_0^{\max}$, where the random phase shifts, the fixed power, IRS-assisted OMA, and NOMA without IRS are used as the benchmarks. As expected, the proposed scheme achieves a remarkable covert rate gain than that achieved by the benchmark schemes as $P_0^{\max}$ and $N$ increase. Specifically, with the same covertness constraint $\varepsilon=0.3$, the covert rate of the proposed scheme with $R_{r,s_{\tilde r}(k)}^{\min}=2$ approaches that with $R_{r,s_{\tilde r}(k)}^{\min}=1$ in the high $P_0^{\max}$ regime, as seen in Fig.~\ref{fig:sim-7}. The main reason is that, as $P_0^{\max}$ becomes sufficiently large, the optimal transmit power of Bob mainly depends on $\bar{\xi}_{w,\text{ul}}^{\ast(-1)}(\varepsilon)$, thus yielding the same covert rate. In addition, as shown in Fig.~\ref{fig:sim-8}, the random phase shift outperforms the fixed power in terms of the covert rate with a large $N$, e.g., $N=120$. The underlying reason is that when $N$ is sufficiently large but $P_r^{\max}$ is finite, the transmit power of Bob is determined by $\frac{P_r^{\max}|g_{ra}|^2-\tilde{\gamma}_\mathrm{th}\sigma_0^2} {\tilde{\gamma}_\mathrm{th}|g_{ba}|^2}$, as indicated by Proposition~4. In this case, the transmit power allocation at Roy and Bob dominates the rate performance.

\begin{figure}[t]
  \centering
  \includegraphics[width=3.0in]{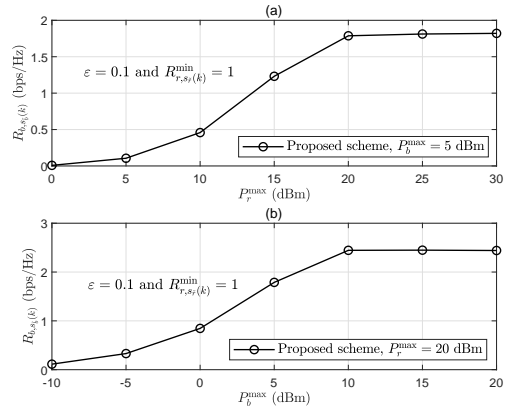}\vspace{-2mm}\\
  \caption{Covert rate versus the individual transmit power budgets at Roy and Bob, where $N=32$.}
  \label{fig:sim-9}\vspace{-3mm}
\end{figure}

Fig.~\ref{fig:sim-9} demonstrates how individual transmit power budgets affect the covert rate of Bob. As seen from Fig.~\ref{fig:sim-9}(a), by fixing $P_b^{\max}$, the covert rate initially increases with $P_r^{\max}$ and then reaches a plateau by further increasing $P_r^{\max}$. This is due to the fact that the transmit power of Bob is determined only by $P_b^{\max}$ with a large $P_r^{\max}$, as implied in Proposition~4. A similar phenomenon can be observed from Fig.~\ref{fig:sim-9}(b) as we fix $P_r^{\max}$ but increase $P_b^{\max}$.

\section{Conclusion}
\label{sec:conclusion}

In this paper, we proposed novel IRS-assisted NOMA schemes for improving the performance of covert wireless communications, where the phase-shift uncertainty of the IRS and the non-orthogonal transmission of Roy are exploited as the new cover medium. Considering the worst-case scenario where Willie can optimally choose his detection threshold, we analytically derived the minimum average detection error probability of Willie to facilitate the performance evaluation. Furthermore, to maximize the covert rates of Bob, the transmit power and the IRS passive beamforming were jointly optimized, subject to the minimum covertness requirement of Willie and the QoS requirement of Roy. Simulation results revealed that the proposed schemes can guarantee covert communications with positive covert rates and significantly outperform other benchmark schemes. Furthermore, it was found that increasing the transmit power for Roy's public signal and the number of reflecting elements at the IRS are beneficial to the covert communication performance.


\section*{Appendix A: Proof of Lemma 1}

Denote $\psi_n=\psi_{n,1}+\psi_{n,2}$, $n=1,\dots,N$, where $\psi_{n,1}=\theta_n^\ast+\arg(h_{an})$ and $\psi_{n,2}=\arg(h_{wn})$. According to \cite{book1}, we know that $\psi_n$, $\psi_{n,1}$, and $\psi_{n,2}$ have a uniform circular distribution. By definition, a circular distribution is a probability distribution of a RV whose values are angles, usually taken to be in the range $[0,2\pi)$. Furthermore, since $\psi_{n,2}$ is uniformly distributed on $[0,2\pi)$, let $\psi_{n,1}=x_1$, we obtain that $\psi_n=x_1+\psi_{n,2}$ is uniformly distributed on $[x_1,x_1+2\pi)=[x_1,2\pi)\cup[2\pi,x_1+2\pi) \overset{(\text{i})}{=}[x_1,2\pi)\cup[0,x_1)=[0,2\pi)$, where step (i) follows from the fact that $\psi_n$ is circularly distributed. Therefore, we prove that $\psi_n$ is uniformly distributed on $[0,2\pi)$.

On the other hand, for $n=1,\dots,N$, we know that $\psi_{n,2}$ is independently distributed while $\psi_{n,1}$ is not, because for different $n$, $\theta_n^\ast$ in $\psi_{n,1}$ is a function that is related to the same $\arg(h_{ab})$ and $\arg(h_{ar})$. Thus, to prove the independence of $\psi_n$, it is necessary to show that $\psi_n$ is independent of $\psi_{n,1}$. To this end, we express the joint PDF of $\psi_n$ and $\psi_{n,1}$ as
\begin{equation}
\label{proof-lemma1-eqn1}
  f_{\psi_{n,1},\psi_n}(x_1,x)=\frac{f_{\psi_{n,1},\psi_{n,2}}(x_1,x_2)} {|\mathbf{J}_{\psi_{n,1},\psi_n}(x_1,x_2)|},
\end{equation}
where $\mathbf{J}_{\psi_{n,1},\psi_n}(x_1,x_2)$ is the Jacobian matrix. Utilizing the fact that $\psi_{n,2}$ is independent of $\psi_{n,1}$ and is uniformly distributed on $[0,2\pi)$, we obtain that $f_{\psi_{n,1},\psi_{n,2}}(x_1,x_2)= f_{\psi_{n,1}}(x_1)f_{\psi_{n,2}}(x_2)=\frac1{2\pi}f_{\psi_{n,1}}(x_1)$. Furthermore, the Jacobian matrix $\mathbf{J}_{\psi_{n,1},\psi_n}(x_1,x_2)$ can be derived as
\begin{equation}
  \mathbf{J}_{\psi_{n,1},\psi_n}(x_1,x_2)=\Bigg[\begin{array}{cc}
    \frac{\partial x_1}{\partial x_1} & \frac{\partial x_1}{\partial x_2}\\
    \frac{\partial x}{\partial x_1} & \frac{\partial x}{\partial x_2}
  \end{array}\Bigg]=\bigg[\begin{array}{cc}
    1 & 0\\
    1 & 1
  \end{array}\bigg],
\end{equation}
such that $|\mathbf{J}_{\psi_{n,1},\psi_n}(x_1,x_2)|=1$. Substituting this result into \eqref{proof-lemma1-eqn1}, we have $f_{\psi_{n,1},\psi_n}(x_1,x)=\frac1{2\pi}f_{\psi_{n,1}}(x_1) =f_{\psi_n}(x)f_{\psi_{n,1}}(x_1)$, where the last equality is based on the fact that $\psi_n$ is uniformly distributed on $[0,2\pi)$. Hence, we conclude that $\psi_n$ is independent of $\psi_{n,1}$.

Summarizing the aforementioned results, Lemma 1 is proved immediately.

\section*{Appendix B: Proof of Theorem 1}

Based on \eqref{DEP-downlink}, we optimize $\xi_{w,\text{dl}}$ in the following three regions of $\tau_\text{dl}$:
\begin{itemize}
  \item If $\tau_\text{dl}<\sigma_0^2+\frac{P_r|h_{aw}|^2}{L(d_{aw})}$, we show that $\xi_{w,\text{dl}}$ is always equal to 1 and cannot be optimized, which is the worst-case detection performance of Willie. Therefore, the optimal $\tau_\text{dl}$ does not exist in this case.

  \item If $\sigma_0^2+\frac{P_r|h_{aw}|^2}{L(d_{aw})}\leq\tau_\text{dl}\leq \sigma_0^2+\frac{(P_r+P_b)|h_{aw}|^2}{L(d_{aw})}$, it is not difficult to verify that $\xi_{w,\text{dl}}$ is a monotonically decreasing function with $\tau_\text{dl}$, and thus, the optimal solution that minimizes $\xi_{w,\text{dl}}$ is $\tau_\text{dl}^\ast=\sigma_0^2+\frac{(P_r+P_b)|h_{aw}|^2}{L(d_{aw})}$.

  \item If $\tau_\text{dl}>\sigma_0^2+\frac{(P_r+P_b)|h_{aw}|^2}{L(d_{aw})}$, we derive the first derivative of $\xi_{w,\text{dl}}$ with respect to $\tau_\text{dl}$ as
      \begin{equation}
      \label{AP-A-eqn1}
        \frac{\partial\xi_{w,\text{dl}}}{\partial\tau_\text{dl}}= \frac{e^{\frac{(P_r+P_b)|h_{aw}|^2/L(d_{aw})+\sigma_0^2-\tau_\text{dl}} {(P_r+P_b)N/(L(d_a)L(d_w))}}}{(P_r+P_b)N/(L(d_a)L(d_w))} -\frac{e^{\frac{P_r|h_{aw}|^2/L(d_{aw})+\sigma_0^2-\tau_\text{dl}} {P_rN/(L(d_a)L(d_w))}}}{P_rN/(L(d_a)L(d_w))}.
      \end{equation}
      By setting $\frac{\partial\xi_{w,\text{dl}}}{\partial\tau_\text{dl}}=0$, the optimal solution is obtained as
      \begin{equation}
      \label{AP-A-eqn2}
        \tau_\text{dl}^\dag=\sigma_0^2+\frac{P_r(P_r+P_b)N}{P_bL(d_a)L(d_w)} \ln\Big(\frac{P_r+P_b}{P_r}\Big).
      \end{equation}
      Using results in \eqref{AP-A-eqn1} and \eqref{AP-A-eqn2}, the monotonicity of $\xi_{w,\text{dl}}$ is discussed as follows: 1) If $\tau_\text{dl}^\dag\geq\sigma_0^2+\frac{(P_r+P_b)|h_{aw}|^2}{L(d_{aw})}$, we can prove that $\frac{\partial\xi_{w,\text{dl}}}{\partial\tau_\text{dl}}<0$ for $\tau_\text{dl}\in\big(\sigma_0^2+\frac{(P_r+P_b)|h_{aw}|^2}{L(d_{aw})}, \tau_\text{dl}^\dag\big)$ while $\frac{\partial\xi_{w,\text{dl}}}{\partial\tau_\text{dl}}>0$ for $\tau_\text{dl}\in\big(\tau_\text{dl}^\dag,+\infty\big)$. This implies that $\xi_{w,\text{dl}}$ first decreases and then increases with respect to $\tau_\text{dl}$, and the optimal solution for minimizing $\xi_{w,\text{dl}}$ is given by $\tau_\text{dl}^\ast=\tau_\text{dl}^\dag$. 2) If $\tau_\text{dl}^\dag<\sigma_0^2+\frac{(P_r+P_b)|h_{aw}|^2}{L(d_{aw})}$, it shows that $\xi_{w,\text{dl}}$ is monotonically increasing with respect to $\tau_\text{dl}$ since $\frac{\partial\xi_{w,\text{dl}}}{\partial\tau_\text{dl}}>0$ for $\tau_\text{dl}\in\big(\sigma_0^2+\frac{(P_r+P_b)|h_{aw}|^2}{L(d_{aw})},+\infty\big)$. In addition, we know that $\xi_{w,\text{dl}}$ is a continuous function of $\tau_\text{dl}$. Thus, the optimal solution is $\tau_\text{dl}^\ast=\sigma_0^2+\frac{(P_r+P_b)|h_{aw}|^2}{L(d_{aw})}$.
\end{itemize}
Summarizing the above results, we prove Theorem 1.

\section*{Appendix C: Proof of Propositions 1 and 2}

We can simply prove Proposition 1 by contradiction. Suppose that the maximum objective value of problem (P2) is achieved when the equality in $P_r+P_b\leq P_{\max}$ does not hold, i.e., $P_r+P_b<P_{\max}$. Then, by fixing $P_r$, we scale up $P_b$ by a factor $\eta$ ($\eta>1$) to make $P_r+P_b=P_{\max}$ hold. In this case, we can always obtain a larger objective value of (P2) since $R_{b,s_b(k)}$ is a monotonically increasing function with respect to $P_b$. This contradicts the original assumption of $P_r+P_b<P_{\max}$.

Next, our attention is shifted to the proof of Proposition 2. Let $P_r=P_{\max}-P_b$ and take the first derivative of $\bar{\xi}_{w,\text{dl}}^\ast$ with respect to $P_b$, we can verify that $\partial\bar{\xi}_{w,\text{dl}}^\ast/\partial P_b<0$. Finally, we complete the proof.

\section*{Appendix D: Proof of Proposition 3}

First, as shown in step 3 of Algorithm~\ref{alg:1}, the optimal solutions $P_r(t+1)$ and $P_b(t+1)$ are derived for given $\mathbf{\Theta}(t)$, and thus, we have the following inequality
\begin{equation}
\label{AP-D-eqn1}
  R_{b,s_b(k)}\big(\mathbf{\Theta}(t),P_b(t),P_r(t)\big)\leq R_{b,s_b(k)}\big(\mathbf{\Theta}(t),P_b(t+1),P_r(t+1)\big).
\end{equation}
Then, as shown in step 4 of Algorithm~\ref{alg:1}, the optimal solution $\mathbf{\Theta}(t+1)$ is obtained for given $P_r(t+1)$ and $P_b(t+1)$, and thus, the following inequality holds
\begin{equation}
\label{AP-D-eqn2}
  R_{b,s_b(k)}\big(\mathbf{\Theta}(t),P_b(t+1),P_r(t+1)\big)\leq R_{b,s_b(k)}\big(\mathbf{\Theta}(t+1),P_b(t+1),P_r(t+1)\big).
\end{equation}
By summarizing \eqref{AP-D-eqn1} and \eqref{AP-D-eqn2}, we can readily obtain that
\begin{equation}
\label{AP-D-eqn3}
  R_{b,s_b(k)}\big(\mathbf{\Theta}(t),P_b(t),P_r(t)\big)\leq R_{b,s_b(k)}\big(\mathbf{\Theta}(t+1),P_b(t+1),P_r(t+1)\big).
\end{equation}
This indicates that the objective value of problem (P1) is non-decreasing over the iterations. On the other hand, the objective value of problem (P1) clearly has an upper bound for any choice of feasible $\mathbf{\Theta}$ and $P_a^{\max}$. Hence, Algorithm~\ref{alg:1} is proved to converge.

\end{document}